\documentclass[12pt]{article} 
\usepackage[sectionbib]{natbib}
\usepackage{array,epsfig,fancyhdr,rotating}
\usepackage[]{hyperref}  
\usepackage{sectsty, secdot}
\usepackage{geometry}
\usepackage{hyperref}
\usepackage{setspace}
\usepackage{comment}
\sectionfont{\fontsize{12}{14pt plus.8pt minus .6pt}\selectfont}
\renewcommand{\theequation}{\thesection\arabic{equation}}
\subsectionfont{\fontsize{12}{14pt plus.8pt minus .6pt}\selectfont}
\usepackage{amsmath}
\usepackage{amssymb}
\usepackage{amsfonts}
\usepackage{multirow}
\usepackage{amsthm}
\usepackage[utf8]{inputenc}
\usepackage{placeins}
\usepackage{graphicx}
\usepackage{enumitem}
\usepackage{authblk}
\usepackage{booktabs, float}
\usepackage{xcolor}
\usepackage{bm}
\usepackage{array}
\newcommand{\red}[1]{{\leavevmode\color{black}#1}}
\newcolumntype{H}{>{\setbox0=\hbox\bgroup}c<{\egroup}@{}}

\DeclareMathOperator{\bX}{\mathbf{X}}
\DeclareMathOperator{\bZ}{\mathbf{Z}}

\DeclareMathOperator{\bS}{\mathbf{S}}
\DeclareMathOperator{\bW}{\mathbf{W}}
\DeclareMathOperator{\bU}{\mathbf{U}}
\DeclareMathOperator{\bD}{\mathbf{D}}
\newcommand{\bV}{\mathbf{V}}
\DeclareMathOperator{\bbeta}{\bm{\beta}}
\DeclareMathOperator{\btheta}{\bm{\theta}}
\DeclareMathOperator{\bgamma}{\bm{\gamma}}
\DeclareMathOperator{\bSigma}{\mathbf{\Sigma}}

\newtheorem{condition}{Condition}

\newtheorem{theorem}{Theorem}
\newtheorem{lemma}{Lemma}

\theoremstyle{definition}

\begin{document}




\title{A Linear Errors-in-Variables Model with Unknown Heteroscedastic Measurement Errors}
\author[1]{Linh H. Nghiem\thanks{Corresponding author: linh.nghiem@sydney.edu.au}} 
\author[2]{Cornelis J. Potgieter} 
\affil[1]{University of Sydney}
\affil[2]{Texas Christian University \& University of Johannesburg}
\date{}
\maketitle
\begin{quotation}
\noindent {\it Abstract:}
    In the classic measurement error framework, covariates are contaminated by independent additive noise. This paper considers parameter estimation in such a linear errors-in-variables model where the unknown measurement error distribution is heteroscedastic across observations. We propose a new generalized method of moment (GMM) estimator that combines a moment correction approach and a phase function-based approach. The former requires distributions to have four finite moments, while the latter relies on covariates having asymmetric distributions. The new estimator is shown to be consistent and asymptotically normal under appropriate regularity conditions. The asymptotic covariance of the estimator is derived, and the estimated standard error is computed using a fast bootstrap procedure. The GMM estimator is demonstrated to have strong finite sample performance in numerical studies, especially when the measurement errors follow non-Gaussian distributions.
\vspace{9pt}

\end{quotation}\par

\def\thefigure{\arabic{figure}}
\def\thetable{\arabic{table}}

\renewcommand{\theequation}{\thesection.\arabic{equation}}

\fontsize{12}{14pt plus.8pt minus .6pt}\selectfont

\clearpage
\section{Introduction}

The errors-in-variables linear model arises when certain covariates suffer from measurement error contamination. This can stem from sources like instrumentation and self-reporting errors, as well as the inadequate use of short-term measurements as proxies for long-term variables. Ignoring measurement error can result in biased \red{estimators}, see \cite{carroll2006measurement} regarding the importance of measurement error correction in understanding the effects of the covariates on the outcome. This paper considers a heteroscedastic measurement error setting, allowing the measurement error covariance to vary across observations. This observation-specific measurement error variance structure, treated as unknown, requires estimation from replicate data. We adopt the classic additive measurement error model wherein the contaminated covariates, i.e the surrogates, are treated as the sum of the true covariates and independent measurement errors, so surrogate variances exceed true covariate variances.

One of the first papers to address the problem of a predictor variable contaminated by measurement error is \cite{wald1940fitting}. Since then, many parametric methods have been proposed, such as the maximum likelihood approach of \cite{higdon2001maximum}. The conditional scores approach \citep{stefanski1987conditional} and the conditional quasi-likelihood approach \citep{hanfelt1997approximate} require the conditional distributions of the outcomes and the contaminated covariates to be specified, both in terms of the true covariates. Regression calibration \citep{carroll1990approximate} estimates the true covariates from the contaminated covariates in a validation sample. Simulation-extrapolation (SIMEX) \citep{stefanski1995simulation} is a computationally-intensive method that adds additional measurement error to estimate model parameters and then extrapolates to the error-free case. All the methods listed in this paragraph require parametric specifications for some distributional components of the model. 

Our paper proposes an efficient \red{distribution-free} estimator for a linear errors-in-variables model with heteroscedastic measurement errors. Our estimator combines two existing methods: a moment correction approach and a phase-function based estimator. The moment correction approach dates back to \cite{reiersol1941confluence} and has also been considered by \cite{gillard2014method} and \cite{erickson2014minimum}. This method requires the existence of model moments up to order $2M$, where $M\geq 2$ is the number of moments that are used to derive moment equations. In contrast, the phase function-based estimator, proposed by \cite{nghiem2020estimation}, does not impose moment conditions on the underlying random variables, but requires the true covariates to have asymmetric distributions. These authors define minimum distance estimators based on the difference of two empirical phase functions. Our paper proposes a new method combining the moment-based and phase function-based approaches within a generalized method of moments (GMM) framework. We assume that the measurement errors have a joint symmetric distribution, but both the distribution type and the observation-specific  covariance matrix are unknown. Our estimator relaxes the asymmetry condition of \cite{nghiem2020estimation} for a common scenario in practice where all observations have at least two replicates \citep{carroll2006measurement}. We propose two different weighting schemes to construct weighted empirical phase functions that account for measurement error heteroscedasticity. Furthermore, we present a computationally efficient bootstrap technique to estimate the covariance matrix, serving both GMM estimator computation and standard error estimation. Simulation studies and a data application demonstrate that our combined GMM estimator have strong finite sample performance.

\red{The remaining sections of the paper are organized as follows. Section 2 reviews moment-corrected and phase function estimators. Section 3 introduces the GMM estimator along with a bootstrap approach for  covariance matrix estimation. Section 4 conducts simulation studies to compare estimation methods and assess standard error recovery. Section 5 presents an illustrative NHANES dataset analysis, and Section 6 concludes the study.}

\section{Moment-corrected and phase function estimators}
\label{sec: MC and Phase Intro}

\subsection{Heteroscedastic errors-in-variables model formulation} \label{sec:ModelFormulation}

\red{Let $\mathcal{X}=\{\mathcal{D}_1,\ldots,\mathcal{D}_n\}$ denote an observed sample following a linear errors-in-variables (EIV) model. Here, $\mathcal{D}_j = (\bW_j^{(n_j)},\tilde{\bZ}_j,y_j)$ is a random sample following EIV model structure that}
\begin{equation}
y_j = \bX_j^\top \bbeta_0 + \bZ_j^\top\bgamma_0 + \varepsilon_j\, \text{ and }\, \bW_{jk} = \bX_{j} + \bU_{jk},
\label{eq: multiple_linear EIV}
\end{equation}
for $k=1,\ldots,n_j$ and $j=1,\ldots,n$. In this model, we let $y_j \in \mathbb{R}$ denote the outcome of interest, $\bX_j =(X_{j1},\ldots,X_{jp})^\top \in \mathbb{R}^p$ denote the true values of the error-prone covariates, $\bZ_j = \big(1,\tilde{\bZ}_j^\top\big)^\top$ with $\tilde{\bZ}_j =(Z_{j1},\ldots,Z_{jq})^\top \in \mathbb{R}^q$ denote the error-free covariates, and $\bW_{jk} = (W_{jk,1},\ldots,W_{jk,p})^\top$ denote a contaminated version of $\bX_j$ subject to independent measurement error $\bU_{jk}=(U_{jk,1},\ldots,U_{jk,p})^\top$.
Furthermore, we let $\bW_j^{(n_j)}=(\bW_{j1},\ldots,\bW_{jn_j})$ and $\bU_j^{(n_j)}=(\bU_{j1},\ldots,\bU_{jn_j})$ denote, respectively, the collections of contaminated replicates and measurement errors associated with the $n_j\geq 2$ replicates of the $j$th observation, and $\varepsilon_j$ denote the usual regression error. Also, $\bbeta_0 = (\beta_{01}, \ldots, \beta_{0p})^\top \in \mathbb{R}^{p}$ and $\bgamma_0 = (\gamma_{00},\gamma_{01}, \ldots, \gamma_{0q})^\top \in \mathbb{R}^{q+1}$ denote, respectively, the coefficients vectors associated with $\bX_j$ and $\bZ_j$. We assume the measurement errors $\bU_{jk}$ have mean zero and covariance $\bSigma_j$ that is potentially distinct for all observations. Consequently,  we have $\mathrm{Var}[\bW_{jk}] = \bSigma_x + \bSigma_j$ where $\bSigma_x = \mathrm{Var}[\bX_j]$ for $k=1,\ldots,n_j$ and $j=1,\ldots,n$. The regression errors $\varepsilon_j$'s are assumed to be independent and identically distributed (\textit{iid}) with mean zero and variance $\sigma_\varepsilon^2$. 

In subsequent sections of this paper, several conditions will be important when considering the estimation methods for the linear EIV model \eqref{eq: multiple_linear EIV}. These conditions are now presented and discussed. To this end, let operator $\perp$ denote the independence of random quantities, let $\left\lVert\bm{A}\right\rVert_{\mathrm{max}} = \max\{|a_{jk}|\}$ denote the element-wise infinity norm of an arbitrary matrix $\bm{A}$, and let $i=\sqrt{-1}$ denote the imaginary unit.

\begin{condition}
For the $j$th observation, $\bU_{jk}{\perp}\bU_{jk'}$ for $k\neq k'$, $k,k'\in \{1,\ldots,n_j\}$. Furthermore, $\bU_{j}^{(n_j)}{\perp}\,(\bX_j,\bZ_j,\varepsilon_j)$. Finally, the random quantities $(\bX_j,\bZ_j, \varepsilon_j)$, $j=1,\ldots,n$ are \textit{iid} copies of random variables $(\bX, \bZ, \varepsilon)$.
\label{condition:independence}
\end{condition}
In addition to requiring independent observations, Condition \ref{condition:independence} requires that the measurement error components across the replicates associated within a given observation are mutually independent. Furthermore, the measurement errors, true covariates, and regression errors are required to be mutually independent as well. An example of a well-studied scenario that would violate this assumption is the Berkson error model wherein the observed predictor $\bW_j$ has a smaller variance than the true predictor $\bX_j$, see \cite{song2021nonparametric} for an overview. Many time-series error models would also violate Condition \ref{condition:independence}. Nevertheless, these settings are outside the scope of the current paper.

\begin{condition}
For the $j$th observation, random quantities $\tilde{\bX}_j = [\bX_j^\top,\tilde{\bZ}_j^\top]^\top$ and $\varepsilon_j$ satisfy $\mathrm{E}\left(\big\lVert\tilde{\bX}_j\tilde{\bX}_j^\top\big\rVert_{\mathrm{max}}^4\right)<\infty$ and $\mathrm{E}(\varepsilon_{j}^{4})<\infty$. Also, for the replicates associated with observation $j$, $\mathrm{E}\left(\left\lVert\bU_{jk}\bU_{jk}^\top\right\rVert_{\mathrm{max}}^4\right)<\infty$, $j=1,\ldots,n$. 
\label{condition:fourfinitemoments}
\end{condition}
While this paper imposes no parametric distributional assumption on the underlying variables in the model, we do require that the covariates, regression errors, and measurement errors have distributions with at least four finite moments. Examples of situations where Condition \ref{condition:fourfinitemoments} would be violated would be if the covariates and/or measurement error terms followed a multivariate $t$ distribution with fewer than $4$ degrees of freedom, or a multivariate stable law with index parameter $\alpha<2$. Condition \ref{condition:fourfinitemoments} is central to the moment-corrected approach of Section \ref{sec: MomCor Estim} having an asymptotic normal distribution. While the phase function-based approach of Section \ref{sec: PhsFunc Estim} is less concerned with the higher-order moments, the following two conditions are central to it.

\begin{condition}
The replicate measurement error vectors $\bU_{jk}$ have a distribution symmetric about zero with strictly positive characteristic function, $\phi_{\bU_j}(\bm{t}) = \mathrm{E}[\exp(i\bm{t}^\top\bU_{jk})]>0,\ k=1,\ldots,n_j$, \red{for all $\bm{t} = (t_1,\ldots,t_p)^\top\in\mathbb{R}^p$ and possibly distinct for each $j=1,\ldots,n$}. Similarly, the regression errors $\varepsilon_j$ have a distribution that is symmetric around zero with strictly positive characteristic function $\phi_\varepsilon(t) = \mathrm{E}[\exp(it\varepsilon_j)]>0$ for all $t\in\mathbb{R}$ and \red{common to all} $j=1,\ldots,n$.
\label{condition:symmetricerror}
\end{condition}
Many distributions commonly encountered in the measurement error literature satisfy Condition \ref{condition:symmetricerror}, including the Gaussian, the Laplace, and Student's $t$ distributions. Excluded by this condition are distributions only taking values on a bounded interval, for example, the (multivariate) uniform distribution, because the corresponding characteristic functions are negative for some $\bm{t}$.

In order to state the next condition, the phase function of a random variable has to be defined. For an arbitrary random variable $V$, let $\phi_V(t)$ denote the characteristic function of $V$. The phase function of $V$ is then defined as $\rho_V(t) = \phi_V(t)/| \phi_V(t) |$ with $| \phi_V(t) |^2 = \phi_V(t)\bar{\phi}_V(t)$ being the squared complex norm and $\bar{\phi}_V(t)$ the complex conjugate of $\phi_V(t)$. For a more in-depth discussion of the phase function and its properties, consult \cite{delaigle2016methodology} and \cite{nghiem2020estimation}.

\begin{condition}
	Let $V(\bbeta,\bgamma) = \bX^\top \bbeta + \bZ^\top \bgamma$ and let $\rho_V(t|\bbeta,\bgamma)$ denote the corresponding phase function of $V$. Note that this phase function depends on parameters $(\bbeta,\bgamma)$. Then, $\rho_V(t|\bbeta,\bgamma)$ is continuously differentiable with respect to all elements of $\bbeta$ and $\bgamma$. Furthermore, $\partial \rho_V(t|\bbeta,\bgamma)/\partial \beta_k \neq 0$ for $k=1,\ldots,p$ and $\partial \rho_V(t|\bbeta,\bgamma)/\partial \gamma_k \neq 0$ for $k=0,\ldots,q$. 
	\label{condition:asymmetric}
\end{condition}

Condition \ref{condition:asymmetric} may appear esoteric. In essence, this condition imposes a joint skewness requirement on the true covariates, as any symmetric variable independent of the other variables will not contribute to the phase function in any way. A related sufficient condition used by \cite{nghiem2020estimation} is that for covariates $\tilde{\bX}=(\bX^\top,\tilde{\bZ}^\top)^\top = (\tilde{X}_1,\ldots,\tilde{X}_{p+q})^\top$ and true parameter values $\tilde{\bbeta}_0 = (\tilde{\beta}_{01},\ldots,\tilde{\beta}_{0,p+q}) = (\beta_{01},\ldots,\beta_{0p},\gamma_{01},\ldots,\gamma_{0q})^\top$, there exists no subset of variables $\mathcal{P} \subseteq \{1,\ldots, p+q\}$ such that $\sum_{k\in \mathcal{P}} \tilde{\beta}_{0k} \tilde{X}_k$ has a symmetric distribution.

\begin{condition}
The true parameter $\btheta_0 = (\bbeta_0^\top, \bgamma_0^\top)^\top$ is an interior point of a compact and convex parameter space $\bm\Theta \subseteq \mathbb{R}^{p+q+1}.$
\label{condition:compact}
\end{condition}
Condition \ref{condition:compact} is a regularity condition imposed on the parameter space; here it is satisfied when all parameter values are finite.

\subsection{Moment correction}
\label{sec: MomCor Estim}

Moment correction is a well-established approach to estimate the parameters of the heteroscedastic EIV model as per equation \eqref{eq: multiple_linear EIV}. As moment correction is also central to the new estimation method proposed in Section \ref{sec:CombinedEstimator}, a brief overview is provided here. The interested reader can consult \citet[Section 5.4]{buonaccorsi2010measurement} for  more details.

Let $\bm{W}_j = n_{j}^{-1} \sum_{k=1}^{n_j} \bm{W}_{jk}= \bm{X}_j + \bm{U}_j$ denote the averaged contaminated replicates of the $j$th observation. Here, $\bm{U}_j = n_{j}^{-1} \sum_{k=1}^{n_j} \bm{U}_{jk}$, whose variance is $\mathrm{Var}\left(\bm{U}_{j}\right)= n_j^{-1}\bm{\Sigma}_j$. Moment correction relies on the corrected $L_2$ norm
$    L(\bbeta,\bgamma) = n^{-1}\sum_{j=1}^{n} \left(y_j -  \bW_j^\top\bbeta - \bZ_j^\top\bgamma   \right)^2 - n^{-1} \sum_{j=1}^{n} n_j^{-1} \bbeta^\top \bSigma_j \bbeta,
    \nonumber
$
which satisfies $\mathrm{E}[L(\bbeta_0,\bgamma_0)] = \sigma_\varepsilon^2$ when  Condition \ref{condition:independence} holds. The gradients corresponding to this \red{function} are
$$
\begin{aligned}
&   \bS_{L,\beta} = \dfrac{\partial L}{\partial \bbeta} = - \dfrac{2}{n}\sum_{j=1}^{n} \bW_j \left(y_j - \bW_j^\top\bbeta - \bZ_j^\top\bgamma   \right) - \dfrac{2}{n}\sum_{j=1}^{n}  \dfrac{1}{n_j}\bSigma_j \bbeta,\\
&    \bS_{L,\gamma} = \dfrac{\partial L}{\partial \bgamma} = -\dfrac{2}{n}\sum_{j=1}^{n} \bZ_j \left(y_j - \bW_j^\top\bbeta - \bZ_j^\top\bgamma   \right).
\end{aligned}
$$
The moment corrected estimator is  subsequently defined to be the solution of the $p+q+1$ estimating equations $\bS_L = [\bS_{L,\beta}^\top,  \bS_{L,\gamma}^\top]^\top = \bm{0}$. These estimating equations are well-defined in a statistical sense whenever the underlying variables have at least two finite moments. However, to establish that the estimators are asymptotically normally distributed, Conditions \ref{condition:fourfinitemoments} and \ref{condition:compact} are required to ensure the  variances of the gradients $\bS_{L, \beta}$ and $\bS_{L,\gamma}$ are finite. 

To implement the above moment correction, knowledge of the variance-covariance matrices $\bSigma_j$ is required. When these are unknown, they can be consistently estimated using 
\begin{equation}
\hat\bSigma_j = \dfrac{1}{n_j(n_j-1)} \sum_{k=1}^{n_j-1}\sum_{k^\prime>k}^{n_j} (\bW_{jk} - \bW_{jk^\prime}) (\bW_{jk} - \bW_{jk^\prime})^\top.
\label{eq: hatSigma_j}
\end{equation}
The covariance estimator in \eqref{eq: hatSigma_j} follows upon noting that $$\mathrm{E}\left[(\bW_{jk} - \bW_{jk^\prime}) (\bW_{jk} - \bW_{jk^\prime})^\top\right] = \mathrm{E} \left[(\bU_{jk} - \bU_{jk^\prime})(\bU_{jk} - \bU_{jk^\prime})^\top  \right] = 2\bSigma_j$$ for any pair of replicates $(\bW_{jk},\bW_{jk'})$ with $k\neq k'$. The second equality follows from the independence of measurement error terms $(\bU_{jk},\bU_{jk^\prime})$, and the assumed mean and covariance structure of the $\bU_{jk}$. Subsequently, the estimator $\hat\bSigma_j$ is defined by averaging over the squared differences of the $n_j(n_j-1)/2$ such pairs associated with the $j$th observation. Therefore, in practice, the gradient $\bS_{L,\beta}$  is  replaced by an approximation $\hat{\bS}_{L,\beta}$ which substitutes  unknown $\bSigma_j$ by $\hat{\bSigma}_j$ for $j=1,\ldots,n$. The moment-corrected estimators are then calculated as the solution of the estimating equation
\begin{equation}
    \hat\bS_L = [\hat\bS_{L,\beta}^\top,  \bS_{L,\gamma}^\top]^\top = \bm{0}. 
    \label{eq: est_eq_corrected}
\end{equation}
These estimators $\hat\bSigma_j$ and the gradient $\hat\bS_L$ will be further used in Section \ref{sec:CombinedEstimator}.

\subsection{Phase function-based estimation}
\label{sec: PhsFunc Estim}

Recently, in the context of a homoscedastic EIV model without replicate data, \cite{nghiem2020estimation} proposed a phase function-based estimator. Their approach, unlike the moment-corrected estimator, does not require estimation of the measurement error covariances, but still leads to a consistent estimator even when the underlying random variables do not have finite variances.
We propose here a new variation of the phase function method that adjusts for heteroscedasticity of the measurement error, which is made possible by the availability of replicates. Furthermore, the asymmetric linear combination assumption of \cite{nghiem2020estimation} is replaced by Condition \ref{condition:asymmetric}. 

For the model defined in \eqref{eq: multiple_linear EIV}, define $V_{0j} = \bX_j^\top\bbeta_0 + \bZ_j^\top\bgamma_0$ so that $y_j = V_{0j} + \varepsilon_j$, $j=1,\ldots,n$. Since $(\bX_j, \bZ_j, \varepsilon_j)$ are independent copies of $(\bX, \bZ)$ by Condition \ref{condition:independence}, the random variable $V_{0j}$ is an independent copy of $V_0 := V(\bbeta_0, \bgamma_0) = \bX^\top\bbeta_0 + \bZ^\top\bgamma_0$ and $y_j$ is an independent copy of $Y = V_0 + \varepsilon$.  Recall that $\phi_Y(t)$ and $\rho_Y(t)$ denote, respectively, the characteristic and phase functions of $Y$, and the same holds for $\phi_{V_0}(t)$ and $\rho_{V_0}(t)$ in terms of $V_0$. From Condition \ref{condition:symmetricerror}, the phase function $\rho_Y(t)$ is given by $$
\rho_Y(t) = \frac{\phi_Y(t)}{|\phi_Y(t)|} \stackrel{(i)}{=} \frac{\phi_{V_0}(t)\phi_\varepsilon(t)}{|\phi_{V_0}(t)\phi_\varepsilon(t)|}\stackrel{(ii)}{=}\frac{\phi_{V_0}(t)}{|\phi_{V_0}(t)|} = \rho_{V_0}(t),
$$ 
where (i) follows from the independence of the regression error as per Condition \ref{condition:independence}, and (ii) follows from Condition \ref{condition:symmetricerror}.  Moreover, recalling that $\bW_j$ and $\bU_j$ are the averaged replicates and measurement error terms, respectively. Since the term $\bU_j^\top\bbeta$ also has a symmetric distribution around zero, a similar argument shows that the phase function of arbitrary linear combination $\tilde{V}_j(\bbeta,\bgamma) = \bW_j^\top \bbeta + \bZ_j^\top \bgamma = \bX_j^\top \bbeta + \bZ_j^\top \bgamma + \bU_j^\top \bbeta$ is the same as the phase function of $V(\bbeta, \bgamma)$,  i.e $\rho_{\tilde{V}_j}(t \vert \bbeta, \bgamma) = \rho_{V}(t \vert \bbeta, \bgamma)$ for all $j=1,\ldots, n$.


Our method now proceed by equating two different empirical phase functions. Firstly, based on the outcomes $y_j$, define empirical phase function
$$
\hat\rho_Y(t) = \frac{\sum_{j=1}^{n} \exp(ity_j)}{\left[\sum_{j=1}^{n}\sum_{k=1}^{n} \exp\left\{it\left(y_j - y_k\right) \right\}\right]^{1/2}},
$$
and based on covariates $(\bW_j,\bZ_j)$, we define the \textit{weighted} empirical phase function (WEPF),
\begin{equation}
\hat{\rho}_V(t|\bbeta,\bgamma) = \dfrac{\sum_{j=1}^{n} q_j \exp\left\{it(\bW_j^\top\bbeta + \bZ_j^\top\bgamma)\right\}}{\left(\sum_{j=1}^{n}\sum_{k=1}^{n} q_j q_k \exp\left[it\left\{(\bW_j - \bW_k)^\top\bbeta + (\bZ_j - \bZ_k)^\top\bgamma \right\} \right]\right)^{1/2}}. \label{eq: WEPF}
\end{equation}
where the weights $\{q_j\}_{j=1}^{n}$ satisfy $q_j \geq 0$ and $\sum_{j=1}^{n} q_j = 1$. The phase function-based estimator is motivated by noting that the population level equivalents $\rho_Y(t)$ and $\rho_V(t|\bbeta,\bgamma)$, under the asymmetry imposed on the covariates by Condition \ref{condition:asymmetric}, are equal if and only if $(\bbeta,\bgamma)=(\bbeta_0,\bgamma_0)$. Thus, the estimator for $(\bbeta_0, \bgamma_0)$ is defined to be the minimizer of the discrepancy
\vspace{-1.2em}
\begin{equation}
    D(\bbeta, \bgamma) = \int_{-\infty}^{\infty} \vert \hat\rho_{Y}(t) - \hat\rho_V(t|\bbeta,\bgamma)  \vert^2 \omega(t) dt, 
    \label{eq: discrepancy}
\end{equation}
where $\omega(t)$ is a weighting function to ensure the integral is finite. Direct minimization of \eqref{eq: discrepancy} is computationally expensive. Following the example of \cite{nghiem2020estimation}, we consider the alternative minimization problem
\begin{eqnarray}
\tilde{D}(\bbeta, \bgamma) &=& \int_{0}^{t^*} \bigg[ 
C_y(t) \sum_{j=1}^{n} q_j\sin\left\{t\left(\bW_j^\top\bbeta + 
\bZ_j^\top\bgamma\right)\right\} \notag \\ 
 && - S_y(t) \sum_{j=1}^{n} q_j \cos \left\{t\left(\bW_j^\top\bbeta +  \bZ_j^\top\bgamma\right)\right\}  \bigg]^2 K_{t^*}(t)dt
\label{eq: mult_phase_func_dist}
\end{eqnarray}
where $C_y(t) = n^{-1} \sum_{j=1}^{n} \cos(t\,y_j)$, $S_y(t) = n^{-1}\sum_{j=1}^{n} \sin(t\,y_j)$, and $K_{t^*}(t)$ is a kernel function that is only non-zero on the interval $[0, t^*]$. Note that the two forms \eqref{eq: discrepancy} and \eqref{eq: mult_phase_func_dist} are equivalent for appropriate choices of $\omega(t)$ and $K_{t^*}(t)$. Following  \cite{delaigle2016methodology}, we use $K_{t^*}(t) = (1- t/t^*)^2$ for numerical implementation with $t^*$ being the largest $t$ such that $\vert \hat\phi_y(t) \vert \leq n^{-1/2}$. For any fixed $(\bbeta, \bgamma)$, by a similar argument to \citet{nghiem2018density}, the WEPF in \eqref{eq: WEPF} is a consistent estimator of $\rho_V(\bbeta, \bgamma)$ for any set of weights having $\max_j q_j = O(n^{-1})$. Therefore, our intended goal is to adjust for measurement error heteroscedasticity through an appropriate choice of weights; we will elaborate on this in Section \ref{section: weights}.

\section{A GMM estimator} \label{sec:CombinedEstimator}

\subsection{Generalized method of moments}

The moment-corrected and phase function-based estimators from Section \ref{sec: MC and Phase Intro} rely on different yet complementary sets of assumptions. On the one hand, moment correction is a least squares approach with a variance correction term, thus using information from the first two underlying model moments. On the other hand, as discussed in \cite{nghiem2020estimation}, the phase function method essentially uses all \textit{odd} moments of the underlying model, provided these moments exist. Hence, a method combining these two approaches will generally make use of more model information, allowing for the possibility of a more efficient estimator. In this paper, we describe how to combine the two methods using a generalized method of moments (GMM) approach, which is widely used to estimate parameters in over-identified systems \citep{hansen1982large}.%


To define the GMM estimator, recall that the phase function-based estimator is found by minimizing \red{function} $\tilde{D}(\bbeta,\bgamma)$ in \eqref{eq: mult_phase_func_dist}. The minimizer can be expressed as the solution of the $p+q+1$ estimating equations $\bS_{\tilde{D}} = [\bS_{\tilde{D},\bm\beta}^\top, \bS_{\tilde{D},\bm\gamma}^\top]^\top = \bm{0}$ where
\begin{equation}
\bS_{\tilde{D},\bbeta} = \frac{\partial \tilde{D}}{\partial \bbeta}\ \mathrm{and}\ \bS_{\tilde{D},\bgamma} = \frac{\partial \tilde{D}}{\partial \bgamma}.
\label{eq:S_D}
\end{equation}
Thus, if we simultaneously consider the estimating equations from moment correction, $\hat\bS_L=\bm{0}$ in \eqref{eq: est_eq_corrected}, and the phase-function based estimators, $\bS_{\tilde{D}}=\bm{0}$ defined above in \eqref{eq:S_D}, we have a system of $2(p+q+1)$ estimating equations in terms of the $p+q+1$ model parameters. More formally, let 
\begin{equation}
\bS := \bS(\bbeta,\bgamma) = [\hat{\bS}_L^\top, {\bS}_D^\top]^\top = [\hat{\bS}_{L,\bm\beta}^\top, \bS_{L,\bm\gamma}^\top, {\bS}_{\tilde{D},\bm\beta}^\top, \bS_{\tilde{D},\bm\gamma}^\top]^\top \label{eq: GEE_estim_eqs}
\end{equation}
denote the vector of $2(p+q+1)$ gradient equations. The system $\bS(\bbeta,\bgamma)=\bm{0}$ is generally over-identified and does not have an exact solution. Suppressing the dependence of $\bS$ on $(\bbeta,\bgamma)$, we define a quadratic form in $\bS$, 
\begin{equation}
Q(\bbeta,\bgamma) = \bS^\top \bm{\Omega}_S^{-1} \bS,
\label{eq:Q}
\end{equation}
where $\bm{\Omega}_S = \big[\mathrm{Var}(\bS)\big]_{(\bbeta,\bgamma)=(\bbeta_0,\bgamma_0)}$ is the $2(p+q+1)\times 2(p+q+1)$ covariance matrix corresponding of the gradient equations $\bS$ evaluated at the true parameter values $(\bbeta_0,\bgamma_0)$. The GMM estimator is then defined to be the minimizer of $Q(\bbeta,\bgamma)$.  Minimizing the GMM \red{discrepancy function} $Q(\bbeta,\bgamma)$ is  equivalent to projecting $\bS$ onto a $(p+q+1)$-dimensional subspace and solving the resulting equations, so the GMM estimator can be thought of as the solution to an optimal linear combination of the estimating equations in $\bS$. 

We next study the asymptotic properties of the proposed GMM estimator. Both the consistency and asymptotic normality of our proposed estimator follow from the properties of the GMM approach under suitable regularity conditions. First, we establish the uniform convergence of $Q(\bbeta, \bgamma)$.
\begin{lemma}
Assume that all random variables in the model \eqref{eq: multiple_linear EIV} have at least two finite moments. Then, for $(\bbeta, \bgamma) \in \bm{\Theta}$ as per Condition \ref{condition:compact}, the function $Q(\bbeta, \bgamma) \stackrel{p}{\rightarrow} Q_0(\bbeta, \bgamma)$ uniformly, where $Q_0(\bbeta,\bgamma) = \bS_0^\top\bm\Omega_S^{-1} \bS_0$ and $\bS_0 := \bS_0(\bbeta,\bgamma) = \lim_{n\rightarrow \infty}\mathrm{E}[\bS(\bbeta,\bgamma)]$ is the limiting expectation of the gradient equations. 
\label{lemma:uniformconvergence}
\end{lemma}
The proof of Lemma \ref{lemma:uniformconvergence} is presented in Section S1 of the Supplementary Materials. The proof relies of verifying sufficient conditions for uniform convergence as per Lemma 2.9 of \cite{newey1994chapter}. The uniform convergence of $Q(\bbeta,\bgamma)$ is an essential step in establishing the consistency of the GMM estimator in our next theorem.
\begin{theorem}
Consider the heteroscedastic linear EIV model defined in \eqref{eq: multiple_linear EIV}. Assume Conditions \ref{condition:independence}, \ref{condition:symmetricerror} and \ref{condition:compact} hold. Also assume that all variables in the model have at least two finite moments. Finally, assume the weights $q_j$ used for constructing the empirical phase function in \eqref{eq: WEPF} satisfy $\max_{j} q_j=O(n^{-1})$. Then, the estimator obtained by minimizing $Q(\bbeta,\bgamma) = \bS^\top \bm{\Omega}_S^{-1} \bS$ is consistent for true value $(\bbeta_0,\bgamma_0)$.  
\label{theorem:consistency}
\end{theorem}
 Theorem \ref{theorem:consistency} follows from Theorem 2.1 of \cite{newey1994chapter} combining the uniform convergence in Lemma \ref{lemma:uniformconvergence} along with establishing that $Q_0(\bbeta, \bgamma)$ has a global minimum at the true parameters ($\bbeta_0, \bgamma_0)$. The proof is presented in Section S2 of the Supplementary Materials.  We further note that for the GMM estimator to be consistent, Condition \ref{condition:asymmetric} (covariate asymmetry) is not a requirement. When it does not hold, some of the components of $\bS$ converge to $0$ for all values of the underlying parameters. However, the limiting quadratic form still has a unique minimum at the true parameter values, because the estimating equations originating with the moment-corrected approach do not rely on asymmetry. With this in mind, we focus the GMM method on a situation where $\partial \rho_V(t|\bbeta,\bgamma) / \partial \beta_k \neq 0$ for \textit{some} $k=1,\ldots,p$ and/or $\partial \rho(t|\bbeta,\bgamma) / \partial \gamma_k \neq 0$ for \textit{some} $k=1,\ldots,q$. When some of these partial derivatives are non-zero, then evaluation of the empirical phase function contributes information to the estimation procedure and it is possible that the efficiency of the estimator is improved. Finally, we establish the asymptotic normality of the proposed estimator. 
\begin{theorem}
Consider the heteroscedastic linear EIV model defined in \eqref{eq: multiple_linear EIV}. Assume Conditions \ref{condition:independence}, \ref{condition:fourfinitemoments}, \ref{condition:symmetricerror},  and \ref{condition:compact} hold. Furthermore, assume the weights $q_j$ used for constructing the empirical phase function in \eqref{eq: WEPF} satisfy $ \max_{j}q_j= O(n^{-1})$. Then, the estimator $(\hat{\bbeta}_{\mathrm{gmm}},\hat{\bgamma}_{\mathrm{gmm}})$ obtained by minimizing $Q(\bbeta,\bgamma) = \bS^\top \bm{\Omega}_S^{-1} \bS$ satisfies
\[
n^{1/2}\, \Big\{ \big(\hat{\bm\beta}_{\mathrm{gmm}}^\top \ \hat{\bm\gamma}_{\mathrm{gmm}}^\top\big)^\top - \big(\bbeta_0^\top \ \bgamma_0^\top \big)^\top \Big\} \sim N \left(\boldsymbol{0},  \left(\bm{P}_1\bm\Omega_S^{-1} \bm{P}_1^{\top} \right) ^{-1} \right) 
\]
where $$\bm{P}_1 = \mathrm{E} \Bigg[\left(\frac{\partial\bS}{\partial\bbeta}^\top, ~\frac{\partial\bS}{\partial\bgamma}^\top\right)\ \Bigg]
$$ 
with the expectations in $\bm{P}_1$ evaluated at the true parameter values $(\bbeta_0,\bgamma_0)$. 
\label{theorem:asymptoticnormality}
\end{theorem}

Theorem \ref{theorem:asymptoticnormality} follows from Theorem 3.4 of \cite{newey1994chapter}. Compared to the required conditions for the consistency established in Theorem \ref{theorem:consistency}, Theorem \ref{theorem:asymptoticnormality} requires stronger moment conditions that all the variables in the model having four finite moments as per Condition \ref{condition:fourfinitemoments}. 
We furthermore note that versions of Theorems \ref{theorem:consistency} and \ref{theorem:asymptoticnormality} still hold if we replace the covariance matrix $\bm\Omega_S$ by any positive definite matrix $\bm\Omega_{\ast}$. Nevertheless, the choice of $\bm\Omega_S$ leads to the most asymptotically efficient estimator, as discussed in Section 5.2 of \cite{newey1994chapter}. 

In practice, the covariance matrix $\bm{\Omega}_S$ is unknown and needs to be replaced by a suitable estimator $\hat{\bm{\Omega}}_S$. \textcolor{black}{As per Section 4 of \cite{newey1994chapter}, using a consistent \red{estimator} of $\bm{\Omega}_S$ leaves the asymptotic distribution of the estimators unchanged.} We therefore propose a bootstrap resampling algorithm in Section \ref{subsection: covariancematrix} to obtain a suitable estimator $\hat{\bm{\Omega}}_S$. First, however, we explore the calculation of weights $q_j$, $j=1,\ldots,n$ in the WEPF to adjust for measurement error heteroscedasticity.

\subsection{Choice of phase function weights}
\label{section: weights}

The weighted phase function in \eqref{eq: WEPF} requires the specification of weights $\{q_j\}_{j=1}^{n}$ with $q_j \geq 0$ and $\sum_{j=1}^{n} q_j = 1$. If the underlying distributions were known, it would be possible to directly minimize an asymptotic variance metric directly related to this estimated phase function -- see \cite{nghiem2018density} for a derivation of the pointwise asymptotic variance of such a WEPF. However, since the underlying distributions are assumed unknown, the direct approach is unavailable to us. We therefore present here two weighting schemes that are easily implemented in practice.

For the first approach, note that the variable $\sum_{j=1}^{n} q_j \bW_j^\top\! \bbeta_0$ is an unbiased estimator of $\mathrm{E}(\bX^\top\! \bbeta_0)$. Therefore, one can consider finding a set of weights that minimize the variance of the above mean estimator. Specifically, we have $\mathrm{Var}\left(\sum_j q_j \bW_j^\top\bbeta_0 \right) = \sum_j q_j^2 \bbeta_0^\top (\bSigma_x + n_j^{-1}\bSigma_j)\bbeta_0$, which is minimized by weights
 $   q_j = {a_j^{-1}}/{\sum_{k=1}^{n}a_k^{-1} }$, where $a_j = \bbeta_0^\top (\bSigma_x + n_j^{-1}\bSigma_j) \bbeta_0$. 
Unfortunately, these weights depend on the unknown true $\bbeta_0$ and are therefore impossible to calculate. We propose the following proxy estimator based on a ``minimax'' argument. Let $\lambda_j$ denote the largest eigenvalue of $\bSigma_x + n_j^{-1}\bSigma_j$. We then have $ a_j \leq \lambda_j \Vert\bbeta_0\Vert^2$ with $\Vert\bbeta_0\Vert^2=\sum_{k=1}^{p} \beta_{0k}^2$ denoting the squared $L_2$ norm. Therefore, we calculate weights based on replacing the $a_j$ by the corresponding upper bounds. Note that using the upper bounds has the potential to \textit{underweight} observations with the large measurement error i.e. their influence is even further mitigated. The proposed minimax weights are given by
\begin{equation}
  \hat{q}^{\mathrm{mm}}_j = \dfrac{\hat\lambda_j^{-1}}{\sum_{k=1}^{n}\hat\lambda_k^{-1}}, j=1,\ldots,n, \label{eq: MM weights}
\end{equation}
where $\hat\lambda_j$ is the largest eigenvalue of $\hat{\bm\Sigma}_x + n_j^{-1}\hat{\bm\Sigma}_j$ with $\hat{\bm\Sigma}_j$ given in \eqref{eq: hatSigma_j} and with $\hat\bSigma_x = (n-1)^{-1}\sum_{i=1}^{n}(\bW_j-\overline\bW)(\bW_j-\overline\bW)^\top - n^{-1}\sum_{j=1}^{n}n_j^{-1}\hat\bSigma_j$, with $\overline\bW = n^{-1}\sum_{j=1}^{n} \bW_j$.

For the second approach, note that for all $j=1,\ldots, n$, we have $\mathrm{E}({\bW}_j) = \mathrm{E}(\bX)$. Therefore, the quantity $\hat{\bm\mu}_q = \sum_{j=1}^{n} q_j {\bW}_j$ is an unbiased estimator of $\mathrm{E}(\bX)$. Recalling that $\mathrm{Var}(\bW_j) = \bm\Sigma_x + n_j^{-1} \bm\Sigma_j$, we define the $L_2$ discrepancy
\begin{equation}
    L(\bm{q}) = \sum_{j=1}^{n}({\bW}_j - \hat{\bm\mu}_q)^\top \left(\bm\Sigma_x + n_j^{-1} \bm\Sigma_j\right)^{-1}({\bW}_j - \hat{\bm\mu}_q)^\top, \label{eq: QL distance}
\end{equation}
with $\bm{q}=(q_1,\ldots,q_n)$. Our second weighting scheme proposes minimizing $L(\bm{q})$ in terms of the weights $\bm{q}$ subject to $q_j \geq 0$ and $\sum_{j=1}^{n} q_j = 1$. We note that minimizing $L(\bm{q})$ is equivalent to maximizing the log-likelihood of the $\{\bW_j\}_{j=1}^{n}$ in terms of $\bm{q}$ assuming each $\bW_j$ follows a multivariate normal distribution. We therefore refer to this approach as the quasi-likelihood weighting scheme. In Section S3 of the Supplementary Materials, we show that the quasi-likelihood weights are found by solving a system of linear equations. Replacing $\bm\Sigma_x$ and the $\bm\Sigma_j$ in \eqref{eq: QL distance} with their corresponding estimator $\hat{\bm\Sigma}_x$ and $\hat{\bm\Sigma}_j$, we denote the resulting minimizing weights by $\hat{q}_j^{\mathrm{ql}}$ for $j=1,\ldots,n$. The performance of the minimax and quasi-likelihood weights are further considered in the simulation studies of Section \ref{sec: Simulations}.

\subsection{GMM covariance matrix and standard error estimation }
\label{subsection: covariancematrix}

Implementation of the proposed GMM estimator requires a suitable \red{estimator} for $\bm\Omega_S$ in \eqref{eq:Q}, where $\bm\Omega_S$ is the covariance matrix of the $2(p+q+1)$ gradient equations $\bS$ at the true parameter values $(\bbeta_0, \bgamma_0)$. We propose a computationally--efficient strategy \red{based on the estimating function bootstrap approach of \citet{hu1997estimating}; this strategy is also used in \citet{nghiem2020estimation} for estimating the variances of the phase function-based estimators.}

\red{Recall that $\mathcal{X}=\{\mathcal{D}_1,\ldots,\mathcal{D}_n\}$ denotes the random sample from the linear EIV model with $\mathcal{D}_j = (\bW_j^{(n_j)},\tilde{\bZ}_j,y_j)$, $j=1,\ldots,n$. The $b$th bootstrap sample, $\mathcal{X}_b^\ast = \{\mathcal{D}_{b1}^\ast,\ldots,\mathcal{D}_{bn}^\ast\}$, $b=1,\ldots,B$, is obtained by sampling $n$ times with replacement from $\mathcal{X}$. No re-sampling is done at the replicate level.} 
For the $b$th bootstrap sample, we compute $\hat{\bSigma}_{b,x}^\ast$, $\hat{\bSigma}_{b,j}^\ast$, and $\hat{q}_{b,j}^\ast$ which correspond to the observation-level measurement error covariance matrices, and the weights for the phase function \red{estimator} calculated using $\mathcal{X}_b$ for $j=1,\ldots,n$. We subsequently use these quantities and our bootstrap sample to evaluate
$\bS_b^\ast(\hat{\bbeta}_{\mathrm{in}},\hat{\bgamma}_{\mathrm{in}}) = [\hat{\bS}_{b,L}^\ast(\hat{\bbeta}_{\mathrm{in}},\hat{\bgamma}_{\mathrm{in}})^\top\ \bS_{b,\tilde{D}}^\ast(\hat{\bbeta}_{\mathrm{in}},\hat{\bgamma}_{\mathrm{in}})^\top]^\top$ as defined in equation \eqref{eq: GEE_estim_eqs}. \red{Here, $(\hat{\bbeta}_{\mathrm{in}},\hat{\bgamma}_{\mathrm{in}})$ denote consistent initial estimators of $(\bbeta_0, \bgamma_0)$; in our implementation, the moment-corrected estimators $(\hat{\bbeta}_{\mathrm{mc}},\hat{\bgamma}_{\mathrm{mc}})$ are used as initial estimators.} Finally, using the $B$ bootstrap samples, we estimate $\bm{\Omega}_S$ by $\hat{\bm{\Omega}}_S^\ast = {B}^{-1}\sum_{b=1}^{B} \bS_b^\ast(\hat{\bbeta}_{\mathrm{in}},\hat{\bgamma}_{\mathrm{in}})\bS_b^\ast(\hat{\bbeta}_{\mathrm{in}},\hat{\bgamma}_{\mathrm{in}})^\top$. 
The method is fast and can be implemented without the need to minimize bootstrap versions of the \red{discrepancy function} $Q(\bbeta,\bgamma)$;  implementation requires only evaluation of the gradient vector for each bootstrap sample prior to calculating $\hat{\bm{\Omega}}_S^\ast$. We have found that $B=100$ bootstrap samples lead to a good performance. The GMM estimator is finally computed by minimizing $\hat{Q}(\bbeta,\bgamma) = \bS^\top \big(\hat{\bm{\Omega}}_S^\ast\big)^{-1}\bS$. 

\red{As pointed out by a referee, our approach is similar to the traditional two--step feasible GMM approach, in which we first obtain a consistent estimator $(\hat{\bbeta}_{\mathrm{mc}},\hat{\bgamma}_{\mathrm{mc}})$ and then use it to estimate the covariance matrix $\bm\Omega_S$ in the GMM discrepancy function. Note that an iterated GMM approach can be pursued, whereby this estimator and a new round of bootstrap samples are used to update the GMM covariance matrix and then minimize the resulting statistic. Asymptotically, this iterated GMM estimator is equivalent to the two-step estimator, although in finite samples, the relative performance between two--step and iterative GMM estimators have been reported to be mixed, see \cite{hansen1996finite}.}

This bootstrap quantity $\hat{\bm{\Omega}}_S^\ast$ can also subsequently be used to estimate standard errors of the GMM estimators. Based on the GMM covariance matrix from Theorem \ref{theorem:asymptoticnormality}, the asymptotic covariance matrix of $(\hat{\bm\beta}_{\text{gmm}}, \hat{\bm\gamma}_{\text{gmm}})$ is estimated by $(\hat{\bm{P}}_1 \hat{\bm{\Omega}}_S^\ast \hat{\bm{P}}_1^\top)^{-1}$. Here, $\hat{\bm{P}}_1$ an empirical counterpart of the expected gradient $\bm{P}_1$ and is evaluated at the GMM \red{estimator} $(\hat{\bm\beta}_{\text{gmm}}, \hat{\bm\gamma}_{\text{gmm}})$. Specifically, the matrix $\hat{\bm{P}}_1$ is given by
$$
\hat{\bm{P}}_1
= \begin{bmatrix}
2n^{-1}\sum_{j=1}^{n}\bW_j\bW_j^\top - 2n^{-1}\sum_{j=1}^{n} \hat{\bm\Sigma}_j & & 2n^{-1}\sum_{j=1}^{n} \bW_j \bZ_j^\top \\[1em]
2n^{-1}\sum_{j=1}^{n} \bZ_j \bW_j^\top & & 2n^{-1} \sum_{j=1}^{n}\bZ_j \bZ_j^\top \\[1em]
\dfrac{\partial{\hat{\bS}}_{\tilde{D}, \bm\beta}}{\partial \bm\beta}\bigg\rvert_{(\hat{\bm\beta}_{\text{gmm}}, \hat{\bm\gamma}_{\text{gmm}})}  & & \dfrac{\partial{\hat{\bS}}_{\tilde{D}, \bm\beta}}{\partial \bm\gamma}\bigg\rvert_{(\hat{\bm\beta}_{\text{gmm}}, \hat{\bm\gamma}_{\text{gmm}})}  \\[1.5em]
\dfrac{\partial{\hat{\bS}}_{\tilde{D}, \bm\gamma}}{\partial \bm\beta}\bigg\rvert_{(\hat{\bm\beta}_{\text{gmm}}, \hat{\bm\gamma}_{\text{gmm}})} & &\dfrac{\partial{\hat{\bS}}_{\tilde{D}, \bm\gamma}}{\partial \bm\gamma}\bigg\rvert_{(\hat{\bm\beta}_{\text{gmm}}, \hat{\bm\gamma}_{\text{gmm}})}
\end{bmatrix}.
$$
For implementation, the partial derivatives corresponding to the phase function criterion are calculated numerically. The estimated standard errors are given by the square root of the diagonal elements of $(\hat{\bm{P}}_1 \hat{\bm{\Omega}}_S^\ast \hat{\bm{P}}_1^\top)^{-1}$. In Section 4.3, we consider the performance of these estimated standard errors. 

\section{Simulation studies} \label{sec: Simulations}

\subsection{Simulation description}\label{sec:Sim Description}

The finite-sample performance of the GMM estimators was evaluated using a simulation study. We considered three different settings for mutiple linear EIV models \eqref{eq: multiple_linear EIV} where each observation has the same number of replicates, $n_j = n_{\mathrm{rep}}$ for $j=1,\ldots,n$, with $n_{\mathrm{rep}}\in\{2, 3\}$. We chose sample sizes $n \in \{250, 500, 1000\}$ and generate $M=500$ samples for each simulation configuration. A univariate linear EIV model was also considered in Section S4.1 of the Supplementary Materials. Details of the simulation settings are given as follows:

(I) \label{sim: multiple_I} A model with $p=2$ error-prone and $q=0$ error-free covariates. The true covariates were generated from a Gaussian copula \citep{xue2000multivariate}, where the two covariates have scaled half-normal marginals with variance $1$, i.e. $X_{jk} \stackrel{iid}{\sim} (1-2/\pi)^{-1/2}\, \big\vert N(0, 1) \big\vert$ for $k=1,\ldots, n_\text{rep}$, and correlation $0.5$. Three distributions were considered for the measurement error vectors $\bU_{jk}$, namely a bivariate normal, a bivariate $t$ with $2.5$ degrees of freedom, and a contaminated bivariate normal $\bU_{jk} \sim 0.9\, N(\boldsymbol{0}, \bm{\Sigma}_j) + 0.1\, N(\boldsymbol{0}, 10^2 \bm{\Sigma}_j)$. These distributions were all scaled to have covariance matrices $\bSigma_j$ for the replicates associated with the $j$th observation, where $\bSigma_j = \bD_j\!\bm{R}\!\bD_j$ with $\bD_j$ the marginal standard deviations and $\bm{R}$ the correlation matrix. We considered two choices of $\bm{R}$, namely the identity matrix and a model with common correlation $\rho=0.5$ between all measurement error components. The diagonal elements of the matrix $\bm{D}_j$ were independently generated from the uniform distribution $\sqrt{n_j} \times U(\sqrt{0.2}, \sqrt{1.5})$ by which the marginal signal-to-noise ratios $\mathrm{Var}(X_{jk})/\mathrm{Var}(U_{jk})$ range from $2/3$ (fairly weak signal) to $5$ (fairly strong signal). The true model coefficients were set to $\bbeta_0 = (1, 0.5)^\top$ and intercept $\gamma_0 = 2$. The regression error $\varepsilon_j$ was generated to match the distribution of the measurement error in each scenario and with constant variance $\sigma_\varepsilon^2 = 0.25$ for $j=1,\ldots, n$.

(II) \label{sim: multiple_II} A model with $p=2$ error-prone and $q=2$ error-free covariates, differing from setting I due to the inclusion of two error-free covariates. The true covariates $(\bX_j, \bZ_j)$ were generated to have scaled half-normal marginals with the joint structure specified by a Gaussian copula with correlation $0.5$ between each pair of predictors. The true model coefficients were $\bbeta_0 = (1, 0.5)^\top$ and $\bgamma_0 = (2, 1, 0.5)^\top$. All other configurations are equivalent to the specifications in setting I. 

(III)\label{sim:multiple_IV} The model as in setting II, but the two error-free covariates were generated to be symmetric having standard normal marginals. The predictor correlation structure was still specified by a Gaussian copula with correlation parameter $0.5$ between each pair of predictor variables. 

Among the three error distributions considering, the $t_{2.5}$ does not have four finite moments. Consequently, asymptotic normality as per Theorem 2 does not hold in this case. However, the consistency requirements are still satisfied. We also note that setting III departs from Condition \ref{condition:asymmetric}, allowing us to explore the effect of symmetric predictors when using the GMM method.

For each generated sample, we computed several different estimators. \red{Firstly, we compute the true ordinary least squares (OLS) estimator which regresses $y_j$ on the true uncontaminated data $(\bX_j,\bZ_j)$, as well as a naive OLS estimator which regresses $y_j$ on $(\bW_j, \bZ_j)$ with ${\bW}_j$ denoting the averaged replicates for the $j$th observations.  }
Next, we computed the moment-corrected estimator with plug-in covariance matrices $\hat{\bm\Sigma}_j$ as per Section \ref{sec: MomCor Estim}. Finally, we computed three versions of the proposed GMM estimator, corresponding to an equal weighting scheme, the minimax weights, and the quasi-likelihood weights as per Section \ref{section: weights}. For the GMM estimators, covariance matrix $\hat{\bm{\Omega}}_S^\ast$ was estimated using $B=100$ bootstrap samples.

A robust performance metric was adopted to evaluate estimator performance. Let $\btheta_0 = (\bbeta_0^\top, \bgamma_0^\top)^\top$ be the true coefficients, and $\hat{\btheta}^{(m)} = (\hat{\bbeta}^{(m)\top}, \hat{\bgamma}^{(m)\top})^\top $ denote the \red{estimators} obtained using one of the outlined approaches in the $m$th generated sample, $m=1,\ldots, M$. To remove outliers, we constructed the $M \times (p+q+1)$ error matrix $\bm{A}$ with rows $\bm{A}_m = (\hat{\btheta}^{(m)} - \btheta)^\top$. Next, we formed a vector of medians $\bm{A}_{\mathrm{med}}$ by taking the median of each column of $\bm{A}$. Then, we computed the Mahalanobis distance between each row of $\bm{A}$ and $\bm{A}_{\mathrm{med}}$ using the robust minimum covariance determinant estimator of \cite{rousseeuw1999fast}. Finally, we removed the rows with Mahalanobis distances larger than the 90th percentile and calculate the robust mean square error matrix $\mathrm{MSE}_\mathrm{rob} = \tilde{\bm{A}}^\top \tilde{\bm{A}}/\tilde{M}$ where $\tilde{\bm{A}}$ denotes the matrix $\bm{A}$ with the outlier rows removed and $\tilde{M}$ is the number of rows in $\tilde{\bm{A}}$. The quantity $\mathrm{MSE}_\mathrm{rob}$ is a robust estimator of $\mathrm{E}[(\hat{\btheta}-\btheta)^\top(\hat{\btheta}-\btheta)]$, and $\mathrm{det}(1000\times \mathrm{MSE}_\mathrm{rob})$ is the reported performance metric. We reported the determinant rather than the trace as it better accounts for unequal variances for and correlations between estimated parameters. Smaller (larger) values of $\mathrm{det}(\mathrm{MSE}_\mathrm{rob})$ indicate better (worse) performance of estimators in a square error loss sense.

\subsection{Simulation results: Parameter recovery}

Tables \ref{tab:table_setting_II_2replicates} and \ref{tab:table_setting_III_2replicates} present results for simulation settings I and II with $n_{\mathrm{rep}} = 2$; results with $n_{\mathrm{rep}} = 3$ and are presented in Section S4.2 of the Supplementary Materials and show similar conclusions. \red{The true OLS estimator stands out with much smaller performance metrics than those of the estimators computed using data with measurement error; note that this true estimator is not available in practice since the true covariates $\bX_i$'s are not observed.  On the other hand, the naive OLS estimator has the worst performance in all the settings, highlighting the deleterious effect of ignoring measurement error as well as the importance of measurement error correction. Next, we compare all the correction estimators.} For the $t_{2.5}$ and contaminated normal errors, all versions of the GMM estimator outperform the moment-corrected estimator, often with significant efficiency gains. For normally distributed errors, the GMM estimators perform competitively compared to moment-correction. That being said, in the normal error scenarios, GMM occasionally performs worse than moment-correction; see Setting II.  When comparing GMM weighting schemes, there is no clear superior weighting scheme, but minimax and quasi-likelihood weights perform better than equal weighting for non-normal errors. 
\begin{table}[t]
\centering
\caption{Setting I performance of \red{uncontaminated OLS (True), naive OLS (Naive)}, moment-corrected (MC), and GMM estimators with equal (Equal), minimax (MM) and quasi-likelihood (QL) weights with $n_{\mathrm{rep}} = 2$ as measured by $\mathrm{det(1000 \times \mathrm{MSE}_\mathrm{rob})}$.}
\vspace{.4cm}
\label{tab:table_setting_II_2replicates}
\resizebox{.8\textwidth}{!}{\begin{tabular}{llrrrrrrr}
  \hline
  \toprule[1.5pt]
$\bm{R}$ & $\bm{U}$ & $n$ &  True & Naive & MC & \multicolumn{3}{c}{GMM} \\  
& & & & & & Equal & MM & QL \\[2pt] 
  \hline
  \addlinespace
$\rho=0$ & Normal & 250 & 0.012 & 9.799 & 1.394 & 1.040 & 1.020 & 1.102 \\ 
   &  & 500 & 0.001 & 2.889 & 0.240 & 0.224 & 0.222 & 0.234 \\ 
   &  & 1000 & 0.000 & 0.668 & 0.026 & 0.025 & 0.027 & 0.030 \\ 
   & $t_{2.5}$ & 250 & 0.018 & 11.413 & 0.816 & 0.576 & 0.525 & 0.593 \\ 
   &  & 500 & 0.001 & 5.128 & 0.134 & 0.100 & 0.075 & 0.077 \\ 
   &  & 1000 & 0.000 & 1.502 & 0.021 & 0.013 & 0.011 & 0.010 \\ 
   & Cont.Normal & 250 & 0.010 & 17.965 & 1.290 & 0.693 & 0.605 & 0.556 \\ 
   &  & 500 & 0.001 & 4.305 & 0.149 & 0.105 & 0.077 & 0.090 \\ 
   &  & 1000 & 0.000 & 1.143 & 0.022 & 0.014 & 0.009 & 0.011 \\[.5em] 
  $\rho=0.5$ & Normal & 250 & 0.008 & 27.999 & 3.155 & 2.698 & 3.399 & 2.953 \\ 
   &  & 500 & 0.001 & 7.365 & 0.355 & 0.362 & 0.378 & 0.373 \\ 
   &  & 1000 & 0.000 & 1.339 & 0.037 & 0.036 & 0.038 & 0.039 \\ 
   & $t_{2.5}$ & 250 & 0.015 & 34.374 & 1.351 & 0.643 & 0.661 & 0.730 \\ 
   &  & 500 & 0.001 & 8.390 & 0.211 & 0.128 & 0.112 & 0.111 \\ 
   &  & 1000 & 0.000 & 2.583 & 0.026 & 0.017 & 0.016 & 0.016 \\ 
   & Cont.Normal & 250 & 0.008 & 45.002 & 2.976 & 1.405 & 1.048 & 1.228 \\ 
   &  & 500 & 0.001 & 10.191 & 0.287 & 0.193 & 0.120 & 0.157 \\ 
   &  & 1000 & 0.000 & 2.331 & 0.034 & 0.024 & 0.015 & 0.021 \\  \bottomrule[1.5pt]
\end{tabular}}
\end{table}

\begin{table}[t]
\centering
\caption{Setting II performance of \red{uncontaminated OLS (True), naive OLS (Naive)}, moment-corrected (MC), and GMM estimators with equal (Equal), minimax (MM) and quasi-likelihood (QL) weights with $n_{\mathrm{rep}} = 2$ as measured by $ \mathrm{det(1000 \times \mathrm{MSE}_\mathrm{rob})}$.}
\vspace{.4cm}
\label{tab:table_setting_III_2replicates}
\resizebox{.9\textwidth}{!}{\begin{tabular}{llrrrrrrr}
  \hline
  \toprule[1.5pt]
$\bm{R}$ & $\bm{U}$ & $n$ & True & Naive & MC & \multicolumn{3}{c}{GMM} \\  
& & & & & & Equal & MM & QL \\[2pt] 
  \hline
  \addlinespace
   
 $\rho=0$& Normal & 250 & 0.001 & 19.898 & 3.427 & 4.418 & 4.472 & 3.763 \\ 
   &  & 500 & 0.000 & 2.053 & 0.145 & 0.161 & 0.203 & 0.157 \\ 
   &  & 1000 & 0.000 & 0.121 & 0.003 & 0.004 & 0.004 & 0.004 \\ 
   & $t_{2.5}$ & 250 & 0.001 & 24.749 & 0.985 & 0.751 & 0.594 & 0.574 \\ 
   &  & 500 & 0.000 & 2.359 & 0.087 & 0.055 & 0.052 & 0.045 \\ 
   &  & 1000 & 0.000 & 0.139 & 0.001 & 0.001 & 0.001 & 0.001 \\ 
   & Cont.Normal & 250 & 0.001 & 13.924 & 1.564 & 1.200 & 1.048 & 0.793 \\ 
   &  & 500 & 0.000 & 0.985 & 0.043 & 0.042 & 0.038 & 0.033 \\ 
   &  & 1000 & 0.000 & 0.063 & 0.002 & 0.002 & 0.001 & 0.001 \\[.5em] 
  $\rho=0.5$ & Normal & 250 & 0.001 & 83.078 & 11.799 & 11.058 & 11.721 & 12.045 \\ 
   &  & 500 & 0.000 & 5.037 & 0.266 & 0.288 & 0.332 & 0.292 \\ 
   &  & 1000 & 0.000 & 0.286 & 0.007 & 0.008 & 0.009 & 0.009 \\ 
   & $t_{2.5}$ & 250 & 0.001 & 81.873 & 5.663 & 3.799 & 3.669 & 4.338 \\ 
   &  & 500 & 0.000 & 4.953 & 0.130 & 0.104 & 0.097 & 0.086 \\ 
   &  & 1000 & 0.000 & 0.329 & 0.003 & 0.003 & 0.003 & 0.002 \\ 
   & Cont.Normal & 250 & 0.001 & 71.486 & 5.833 & 3.522 & 2.922 & 2.260 \\ 
   &  & 500 & 0.000 & 3.010 & 0.150 & 0.126 & 0.105 & 0.088 \\ 
   &  & 1000 & 0.000 & 0.230 & 0.004 & 0.004 & 0.003 & 0.002 \\
   \bottomrule[1.5pt]
    \end{tabular}}
\end{table}

Table \ref{tab:table_setting_IV_2replicates} presents the results for setting III where the asymmetry condition C4 is violated, again for the case with $n_{\mathrm{rep}}=2$ replicates. The case with $n_{\text{rep}} = 3$ replicates is summarized in Section S4.2 in the Supplementary Materials. Similar to other settings, there is little difference in the performance between the moment-corrected estimator and the GMM estimator when the measurement error is Gaussian. However, we continue to note that GMM represents a substantive improvement over moment correction for contaminated normal and $t_{2.5}$ error distributions. Also, in the latter two error settings, minimax and quasi-likelihood weights outperform equal weighting; neither of the latter two weighting scheme is clearly preferred. 
\begin{table}[t]
\centering
\caption{Setting III performance of \red{uncontaminated OLS (True), naive OLS (Naive)}, moment-corrected (MC), and GMM estimators with equal (Equal), minimax (MM) and quasi-likelihood (QL) weights with $n_{\mathrm{rep}} = 2$ as measured by $\mathrm{det(1000\times\mathrm{MSE}_\mathrm{rob})}$.}
\vspace{.4cm}
\label{tab:table_setting_IV_2replicates}
\resizebox{.85\textwidth}{!}{\begin{tabular}{ll rrrrrrr}
  \hline
  \toprule[1.5pt]
$\bm{R}$ & $\bm{U}$ & $n$ & True & Naive & MC & \multicolumn{3}{c}{GMM} \\  
& & & & & & Equal & MM & QL  \\[2pt] 
  \hline
  \addlinespace
 $\rho=0$ & Normal & 250 & 0.001 & 10.963 & 1.625 & 1.623 & 1.615 & 1.623 \\ 
   &  & 500 & 0.000 & 0.926 & 0.069 & 0.069 & 0.068 & 0.068 \\ 
   &  & 1000 & 0.000 & 0.064 & 0.002 & 0.002 & 0.002 & 0.002 \\ 
   & $t_{2.5}$ & 250 & 0.001 & 11.607 & 0.944 & 0.936 & 0.855 & 0.890 \\ 
   &  & 500 & 0.000 & 0.996 & 0.041 & 0.040 & 0.032 & 0.036 \\ 
   &  & 1000 & 0.000 & 0.062 & 0.001 & 0.001 & 0.001 & 0.001 \\ 
   & Cont.Normal & 250 & 0.001 & 12.622 & 1.665 & 1.663 & 1.467 & 1.572 \\ 
   &  & 500 & 0.000 & 1.477 & 0.068 & 0.070 & 0.066 & 0.067 \\ 
   &  & 1000 & 0.000 & 0.072 & 0.002 & 0.002 & 0.001 & 0.001 \\[.5em] 
  $\rho=0.5$ & Normal & 250 & 0.001 & 54.053 & 7.288 & 7.162 & 7.214 & 7.338 \\ 
   &  & 500 & 0.000 & 2.879 & 0.147 & 0.145 & 0.147 & 0.148 \\ 
   &  & 1000 & 0.000 & 0.169 & 0.004 & 0.004 & 0.004 & 0.004 \\ 
   & $t_{2.5}$ & 250 & 0.001 & 41.193 & 2.852 & 2.782 & 2.374 & 2.524 \\ 
   &  & 500 & 0.000 & 2.484 & 0.078 & 0.074 & 0.067 & 0.070 \\ 
   &  & 1000 & 0.000 & 0.162 & 0.002 & 0.002 & 0.002 & 0.002 \\ 
   & Cont.Normal & 250 & 0.001 & 65.616 & 5.556 & 5.552 & 4.945 & 5.375 \\ 
   &  & 500 & 0.000 & 3.563 & 0.144 & 0.143 & 0.130 & 0.140 \\ 
   &  & 1000 & 0.000 & 0.205 & 0.004 & 0.004 & 0.004 & 0.004 \\ 
    \bottomrule[1.5pt]
\end{tabular}}
\end{table}

We also note that across settings I through III, it is observed that the MSE metric for the naive estimators also decreases as sample size increases. This is an artifact of the scaling used. The interested reader can easily verify that the MSE metric for the naive estimator decreases at a much slower rate than the same metric for the MC and GMM estimators.

In conclusion, the GMM estimator has  a competitive performance compared to the moment-corrected estimator when the measurement errors are Gaussian. However, the relative decrease in $\mathrm{det}(\mathrm{MSE}_{\mathrm{rob}})$ is often so large for the heavier-tailed error distributions that some would be willing to risk a small loss in efficiency were the error distribution closer to a true normal. Furthermore, the two weighting schemes that account for heteroscedasticity tend to result in  better estimators than the equal weighting. 

\subsection{Simulation results: Standard error estimation}
We also performed a simulation study to examine the performance of the asymptotic covariance estimator in estimating the standard errors of the GMM \red{estimators}. The data were simulated from Settings I and III with the sample size $n \in \{500, 1000\}$, the measurement error $\bU_{jk}$ following either a bivariate normal or bivariate $t$ distribution with 2.5 degrees of freedom, and the correlation matrix $\bm{R}$ having $\rho = 0.5$. We reported the \red{Monte Carlo estimates of the true standard errors obtained from $500$ simulated pairs $(\hat{\bm\beta}_{\text{gmm}}, \hat{\bm\gamma}_{\text{gmm}})$ obtained from independently generated datasets, as well as} the average \red{bootstrap plug-in} standard errors \red{as defined in Section \ref{subsection: covariancematrix}} for the GMM \red{estimators} with the minimax weighting scheme $\hat{q}_j^{\text{mm}}$ defined in \eqref{eq: MM weights}, while the results for other weighting schemes are similar and hence are omitted. Table \ref{tab:se_estimation} presents the results for $n_{\text{rep}}=2$, while the results for $n_{\text{rep}}=3$ are presented in Section S4.2 the Supplementary Materials. The average \red{bootstrap plug-in} standard errors are similar to the \red{Monte Carlo} standard error in all the considered settings, suggesting that the bootstrap procedure in Section \ref{subsection: covariancematrix} provides a \red{reliable estimator} for the standard errors of the proposed GMM estimators. 
\begin{table}[ht]
\centering
\caption{\red{Monte Carlo standard errors (MC-SE)} and average of the \red{bootstrap plug-in standard errors (Avg-SE)} for the GMM \red{estimators} with the minimax weighting scheme in simulation settings I and III with $n_\text{rep}=2$ replicates and $\rho = 0.5$.}
\label{tab:se_estimation}
\vspace{.4cm}
\resizebox{.8\textwidth}{!}{
\begin{tabular}{lllrrrr}
   \hline
  \toprule[1.5pt]
Setting & $\bU$ & Coeff & \multicolumn{2}{c}{$n=500$} & \multicolumn{2}{c}{$n=1000$} \\
 & & & \red{MC-SE} & \red{Avg-SE} & \red{MC-SE} & \red{Avg-SE} \\ 
  \hline
  \addlinespace
I & Normal & $\hat\beta_1$ & 0.031 & 0.030 & 0.025 & 0.022 \\ 
   &  & $\hat\beta_2$ & 0.031 & 0.029 & 0.023 & 0.021 \\ 
   &  & $\hat\gamma_{00}$ & 0.043 & 0.044 & 0.033 & 0.032 \\ 
   & $t_{2.5}$ & $\hat\beta_1$ & 0.029 & 0.026 & 0.021 & 0.018 \\ 
   &  & $\hat\beta_2$ & 0.029 & 0.026 & 0.019 & 0.019 \\ 
   &  & $\hat\gamma_{00}$ & 0.038 & 0.036 & 0.028 & 0.025 \\ [1em]
 III & Normal & $\hat\beta_1$ & 0.037 & 0.033 & 0.021 & 0.024 \\ 
   &  & $\hat\beta_2$ & 0.031 & 0.031 & 0.024 & 0.022 \\ 
   &  & $\hat\gamma_{00}$ & 0.055 & 0.050 & 0.038 & 0.036 \\ 
   &  & $\hat\gamma_{1}$ & 0.028 & 0.027 & 0.019 & 0.020 \\ 
   &  & $\hat\gamma_{2}$ & 0.029 & 0.028 & 0.019 & 0.019 \\ 
   & $t_{2.5}$ & $\hat\beta_1$ & 0.030 & 0.028 & 0.018 & 0.021 \\ 
   &  & $\hat\beta_2$ & 0.029 & 0.028 & 0.019 & 0.021 \\ 
   &  & $\hat\gamma_{00}$ & 0.042 & 0.043 & 0.029 & 0.032 \\ 
   &  & $\hat\gamma_{1}$ & 0.024 & 0.026 & 0.019 & 0.018 \\ 
   &  & $\hat\gamma_{2}$ & 0.025 & 0.024 & 0.016 & 0.018 \\ 
   \bottomrule[1.5pt]
\end{tabular}}
\end{table}

\section{Analysis of NHANES data}
The National Health and Nutrition Examination Survey (NHANES) is a long-running research survey conducted by the National Center for Health Statistics (NCHS). The goal of this longitudinal survey study is to assess the health and nutritional status of both adults and children in the United States, tracking the evolution of this status over time. During the 2009-2010 survey period, participants were interviewed and asked to provide their demographic background as well as information about nutrition habits. Participants also undertook a series of health examinations. To assess the nutritional habits of participants, dietary data were collected using two 24-hour recall interviews wherein the participants self-reported the consumed amount for a series of food items during the 24 hours prior to each interview. Based on these recalls, daily aggregated consumption of water, food energy, and other nutrition components such as total fat and total sugar consumption were computed. We used the 2009-2010 NHANES dietary data to illustrate our new GMM estimation procedure.

In this illustrative analysis, we considered the relationship between participants' BMI (outcome of interest) and their age as well as daily aggregates of energy, protein and fat consumption. As these nutritional variables were calculated based on self-reported data, they are well-known to be subject to measurement error. We restricted our analysis to $n=1595$ white women and treat the nutritional data from each of the interviews as $n_{\mathrm{rep}} = 2$ independently observed replicates. We fitted the multiple linear EIV model to this data, treating the nutritional quantities energy, protein, and fat consumption as error-prone covariates with age considered an error-free covariate. Furthermore, all of these covariates were standardized in the analysis. We computed the naive, moment-corrected, and GMM estimators, the latter with the three different weighting schemes. We further reported the estimated standard errors for these \red{estimators}: The estimated standard errors for the naive \red{estimators} were obtained from the linear regression model of the outcome on all the observed covariates, while the standard errors for the moment corrected estimators were calculated from the robust \red{estimator} given in \citet[Section 5.4]{buonaccorsi2010measurement}. The estimated standard errors for the GMM \red{estimators} were calculated in the same manner as in our simulation using the asymptotic covariance and the bootstrap procedure.

\begin{table}[t]
\centering
\caption{NHANES study estimated coefficients for naive, moment-corrected (MC) and GMM estimators with equal (Equal), minimax (MM), and quasi-likelihood weights (QL) weights.}
\label{tab: data_analysis}
\vspace{.4cm}
\resizebox{.9\textwidth}{!}{\begin{tabular}{lrrrrr}
\hline
  \toprule[1.5pt]
& Naive & MC & \multicolumn{3}{c}{GMM} \\  
&  &  & Equal & MM & QL \\ 
  \hline
  \addlinespace
  Intercept & 26.39 (0.18) & 26.39 (0.18) & 26.63 (0.18) & 26.59 (0.18) & 26.56 (0.18) \\ 
  Energy & -0.86 (0.47) & -2.3 (1.05) & -2.7 (0.78) & -2.59 (0.85) & -2.49 (0.83) \\ 
  Protein & 1.29 (0.34) & 3.35 (1.03) & 3.56 (0.43) & 3.68 (0.49) & 3.54 (0.47) \\ 
  Fat & 0.54 (0.43) & 1.03 (0.99) & 0.88 (0.57) & 0.89 (0.64) & 0.89 (0.59) \\ 
  Age & 3.92 (0.18) & 3.71 (0.19) & 3.82 (0.16) & 3.78 (0.16) & 3.79 (0.16) \\ 
   \bottomrule[1.5pt]
\end{tabular}}
\end{table}
Table \ref{tab: data_analysis} demonstrates that the consequences of ignoring measurement error are apparent -- the naive estimator exhibits dramatic attenuation for all the nutritional variables. On the other hand, the moment-corrected and GMM approaches all appear to correct for the bias as the estimated coefficients for energy, protein and fat intake are much larger in absolute terms. Comparing the different corrected estimators, we note the magnitude of the coefficients corresponding to energy and protein are smaller for moment correction than for GMM. The reverse holds for the coefficient corresponding to fat consumption. The effect of age on BMI is similar across all the estimators, which can be explained by the low correlation between age and each of the nutritional variables (the correlations between age and the averaged energy, protein, and fat intakes are -0.02, 0.10, and 0.03, respectively). Finally, the GMM \red{estimators} tend to have lower estimated standard errors than the moment-corrected \red{estimators}. This possibly reflects information contributions from the phase function beyond the first two moments.

\section{Conclusion}

In this paper, we have explored \red{distribution-free} solutions for parameter estimation in the linear errors-in-variables model with heteroscedastic measurement errors across cases. The newly proposed solution combines the popular moment-corrected estimator with a phase function-based estimator using a GMM framework. On the one hand, the proposed GMM estimator inherits estimating equations from the moment-corrected estimator and in this way is able to relax a strict asymmetry condition imposed on the covariates as in the phase function-based estimator proposed by \cite{nghiem2020estimation}. On the other hand, the proposed GMM estimator inherits estimating equations from the phase function-based estimator that leverages the skewness of the true covariates if present, and introduces observation-specific weighting to account for the measurement error heteroscedasticity. Our simulation studies show that when the measurement errors are normal, the GMM estimator is competitive with the moment-corrected estimator. Nevertheless, when measurement errors are non-normal, the GMM estimator has superior performance across all simulation settings considered, including ones where some covariates are symmetrically distributed. 

We close by noting some future research that could be explored relating to this problem. Firstly, in the multivariate setting, the estimation of observation-level measurement error covariance matrices is a challenge. Some form of covariance regularization could be applied to further improve the estimation efficiency of the regression parameters. \red{Secondly, the idea of a continuously-updating GMM treating the matrix $\bm{\Omega}_S$ as an explicit function of $(\bbeta,\bgamma)$ can be explored. This could further improve estimator performance, but does increase the computational complexity of the problem.} Finally, the development of tools for exploring and quantifying the skewness of the true covariates subject to symmetric measurement error will help practitioners understand when the GMM approach is beneficial.

\newpage
\appendix
\renewcommand{\thesection}{S\arabic{section}}

\centerline{\Large\bf Supplementary Materials}
 \vspace{2pt}

\section{Proof of Lemma 1} \label{sec: Uniform Convg}

In this section, we provide proof of Lemma 1 from Section 3.1 in the main paper. For clarity, the lemma is restated here. To this end, let $\btheta = (\bbeta,\bgamma)$ and recall that $\bS(\btheta) = [\bS_L^\top(\btheta),\bS_{\tilde{D}}^\top(\btheta)]^\top$ denotes the vector of gradient equations with $\bS_L(\btheta)$ the gradient vector of the corrected $L_2$ norm $L(\btheta)$ as defined in Section 2.2, and with $\bS_{\tilde{D}}(\btheta)$ the gradient of the phase function-based statistic $\tilde{D}(\btheta)$ as defined in Section 2.3 of the main paper. Let  $\bS_0(\btheta) = \lim_{n\rightarrow\infty}\mathrm{E}[\bS(\btheta)]$ denote the limiting expectation of the gradient equations. Subsequently, define $$Q_0(\btheta) = \bS_0^\top(\btheta) \bm\Omega_S^{-1} \bS_0(\btheta).$$
Lemma 1 now follows.

\begin{lemma}
Assume that all variables in the model have at least two finite moments. For $\btheta \in \bm{\Theta}\subseteq \mathbb{R}^{p+q+1}$, the function $Q(\btheta) \stackrel{p}{\rightarrow} Q_0(\btheta)$ uniformly. 
\end{lemma}

The proof of Lemma 1 relies on establishing the conditions established in Lemma 2.9 of \citet{newey1994chapter}. Specifically, the proof first shows that function $Q$ converges in probability to $Q_0$ for a fixed value of $\btheta$ and is continuously differentiable for all $\btheta\in \bm{\Theta}$. Then, it establishes a Lipschitz condition that bounds the difference between $Q$ at two arbitrary parameter values. This Lipschitz condition is crucial to show that $Q$ converges uniformly in probability to $Q_0$. Finally, the proof shows that $\bS(\btheta)$ and $\nabla \bS(\btheta) = \partial \bS(\btheta) / \partial \btheta$ converge uniformly to their limiting expected values. Thus, the result from Newey \& McFadden applies and the required uniform convergence of $Q$ to $Q_0$ follows.\bigskip

\noindent\textbf{Proof of Lemma 1.} Consider the function $Q(\btheta) = \bS^\top (\btheta)\bm\Omega_S^{-1} \bS(\btheta)$ and a fixed value $\btheta\in\bm{\Theta}$. Then, $Q(\btheta) \stackrel{P}{\rightarrow} Q_0(\btheta)$ by Slutsky's theorem and the continuous mapping theorem. Moreover, $Q(\btheta)$ is continuously differentiable for all $\btheta \in \bm{\Theta}$. Thus, by the mean value theorem, for $\bm\theta_1,\bm\theta_2\in\bm{\Theta}$, we have
$$Q(\bm\theta_1) - Q(\bm\theta_2) = \nabla Q(\bm\theta_z) (\bm\theta_1-\bm\theta_2),$$
where $\nabla Q(\bm\theta)=\partial Q / \partial \btheta$ and $\bm\theta_z$ is a linear interpolant of $\bm\theta_1$ and $\bm\theta_2$. Applying the Cauchy-Schwarz inequality, we obtain
$$| Q(\bm\theta_1) - Q(\bm\theta_2) | \leq \Vert \nabla Q(\bm\theta_z) \Vert \Vert \bm\theta_1-\bm\theta_2 \Vert.$$

Subsequently, uniform convergence in probability will follow by by Lemma 2.9 of \cite{newey1994chapter} if we can establish the the Lipschitz condition that for some constants $\alpha >0$ and a sequence $B_n = O_p(1)$ such that for $\bm\theta_1$ and $\bm\theta_2$, we have
\begin{equation}
\vert Q(\bm\theta_1) - Q(\bm\theta_2) \vert \leq B_n \Vert \bm\theta_1 - \bm\theta_2 \Vert^\alpha.
\label{lipschitzcondition}
\end{equation}

Observe that equation \eqref{lipschitzcondition} will hold for $\alpha=1$ if $\Vert \nabla Q(\bm\theta_z) \Vert = O_p(1)$. Now, by definition $\nabla Q(\btheta) = 2[\nabla \bS(\btheta)]^\top \bm\Omega_{S}^{-1} \bS(\btheta)$ with $\nabla \bS(\btheta) = \partial \bS(\btheta) / \partial \btheta$ a $(p+q+1)\times(p+q+1)$ matrix of partial derivatives. By another application of the Cauchy-Schwarz inequality, we obtain
$$
\Vert \nabla Q(\btheta_z) \Vert \leq c_S \Vert \nabla \bS(\btheta_z) \Vert \Vert \bS(\btheta_z) \Vert \leq c_S \sup_{\btheta \in \bm\Theta} \Vert \nabla \bS(\btheta) \Vert \cdot \sup_{\btheta \in \bm\Theta} \Vert \bS(\btheta) \Vert
$$ where $c_S$ is a constant depending only on $\bm\Omega_S$ and not on the arguments $\btheta$. Thus, it remains only to be shown that $\bS(\btheta)$ and $\nabla \bS(\btheta)$ converge uniformly to $\mathrm{E}[\bS(\btheta)]$ and $\mathrm{E}[\nabla\bS(\btheta)]$. 

To this end, consider the components $\bS_L(\btheta)$ and $\nabla \bS_L(\btheta)$. These functions are continuous for all $\btheta \in \bm\Theta$. Furthermore, provided the random variables $(\bX_j,\bZ_j)$, $\bU_j$, and $\varepsilon_j$ used to define $Q$ have finite variances, there exists dominating functions $d_{L,1}(\btheta)$ and $d_{L,2}(\btheta)$ such that $\Vert \bS_L(\btheta) \Vert \leq d_{L,1}(\btheta)$ and $\Vert \nabla \bS_L(\btheta) \Vert \leq d_{L,2}(\btheta)$ for all $\btheta \in \bm\Theta$ by a uniform law of large numbers as in \cite{hoadley1971asymptotic} or \cite{potscher1989uniform}.

Next, consider the component $\bS_{\tilde{D}}(\btheta)$ and $\nabla \bS_{\tilde{D}}(\btheta)$. Again, these functions are continuous for all $\btheta \in \bm\Theta$. While slightly tedious, one can also verify that a dominating function $d_{\tilde{D},1}(\btheta)$ exists for $\bS_{\tilde{D}}(\btheta)$ provided $(\bX_j,\bZ_j)$ and $\bU_j$ have finite first moments. Similarly, a dominating function $d_{\tilde{D},2}(\btheta)$ exists for $\nabla \bS_{\tilde{D}}(\btheta)$ provided $(\bX_j,\bZ_j)$ and $\bU_j$ have finite second moments. Again, by the same uniform law of large numbers, uniform convergence is achieved. Consequently, we have $\sup_{\btheta \in \bm\Theta} \Vert \bS(\btheta) \Vert = O_p(1)$ and $\sup_{\btheta \in \bm\Theta} \Vert \nabla \bS(\btheta) \Vert = O_p(1)$, the conditions of \cite{newey1994chapter} are satisfied, and the required uniform convergence of $Q(\btheta)$ to $Q_0(\btheta)$ follows. \qed

\newpage

\section{Proof of Theorem 1} \label{Sec: Consistency Proof}

As $Q_0(\btheta)$ is a positive definite quadratic form in terms of $\bS_0(\btheta)$, the global minimum occurs at a point $\btheta^\ast$ if and only if $\bS_0(\btheta^\ast)=\bm{0}$ for a unique value of $\btheta^\ast$. The proof of Theorem 1 thus relies on showing that $Q_0(\btheta)$, the uniform-in-probability limit of $Q(\btheta)$, has a unique global minimum at the true parameter values $\btheta_0$ due to $\bS_0(\btheta)=\bm{0}$ only when $\btheta = \btheta_0$. Throughout the proof, for any random variable $V$, we let $\phi_V(t)$ denote its characteristic function. For any complex number $z$, let $\text{Re}(z)$ and $\text{Im}(z)$ be its real and imaginary parts, respectively. 
\bigskip

\noindent\textbf{Proof of Theorem 1.} By Lemma 1, the GMM objective function $Q(\btheta)$ converges uniformly in probability to $Q_0(\btheta)$. For consistency of the estimators obtained by minimizing $Q(\btheta)$, it suffices to establish that limiting function $Q_0(\btheta)$ has a unique global minimum at the true parameter $\btheta_0 = (\bbeta_0, \bgamma_0)$. This proof will separately consider the $L_2$ norm and phase function components contributing to $Q(\btheta)$. Particularly, we will prove the two following statements:

\underline{Statement 1}: The first $p+q+1$ elements of $\bm{S}_0(\btheta)$, which correspond to the estimating equations from the corrected $L_2$ norm, has a \textit{unique} solution at $\bm\theta_0$.  

\underline{Statement 2}:  $\bm\theta_0$ is always a solution to the last $p+q+1$ elements of $\bm{S}_0(\btheta)$, which correspond to the  estimating equations from the phase function distance $\tilde{D}$.

Given the two previous statements, Theorem 1 will follow immediately. Indeed, these two statements imply that $\bm\theta_0$ is the unique solution for $\bm{S}_0(\btheta) = 0 $ as a whole. This also establishes that $Q_0(\btheta_0)=0$, meaning $Q_0(\btheta)$ has a unique global minimum of zero at $\btheta_0$. As a result, the estimator $\hat\btheta$ that minimizes $Q(\btheta)$ is consistent for $\btheta_0$. 

Hence it remains to prove the two statements, and we will do it in two separate subsections.  

\subsection{Proof of Statement 1}
Consider first the corrected $L_2$ norm function with estimating equations $\bS_L(\btheta) = [\bS_{L,\bbeta}(\btheta)^\top,\bS_{L,\bgamma}(\btheta)^\top]^\top$ as defined in equation (2.3) of the main paper with
$$
\begin{aligned}
&\bS_{L,\bbeta}(\btheta) = -\frac{2}{n}\sum_{j=1}^{n} \bW_j \big(y_j - \bW_j^\top \bbeta - \bZ_j^\top \bgamma \big) - \frac{2}{n}\sum_{j=1}^{n} \frac{1}{n_j}\bSigma_j \bbeta, \\ 
&\bS_{L,\bgamma} (\btheta) = -\frac{2}{n}\sum_{j=1}^{n} \bZ_j \big(y_j - \bW_j^\top \bbeta - \bZ_j^\top \bgamma \big).
\end{aligned}
$$
The expected values $\mathrm{E}[\bS_{L,\bbeta}(\btheta)]$ and $\mathrm{E}[\bS_{L,\bgamma}(\btheta)]$ are found by evaluating the conditional expectations of the component summands. For $\bS_{L,\bbeta}(\btheta)$, we have
$$\begin{aligned}
    &\ \mathrm{E}\Big[\bW_j \big(y_j - \bW_j^\top \bbeta - \bZ_j^\top \bgamma \big)\Big|\bX_j,\bZ_j\Big] \\
    =&\ \mathrm{E}\Big[ \big(\bX_j+\bU_j\big) \big(\bX_j^\top \bbeta_0 + \bZ_j^\top \bgamma_0 + \varepsilon_j - \bX_j^\top \bbeta - \bZ_j^\top \bgamma - \bU_j^\top \bbeta\big)\Big|\bX_j,\bZ_j\Big] \\
    \stackrel{(i)}{=}&\ \bX_j \bX_j^\top \big(\bbeta_0 - \bbeta \big) + \bX_j \bZ_j^\top \big(\bgamma_0 - \bgamma \big) - n_j^{-1}\bSigma_j \bbeta,
\end{aligned}$$
where step $(i)$ follows from the independence of $\bU_j$ and $(\bX_j,\bZ_j)$, as well as noting that $\mathrm{E}\big[ \bU_j \bU_j^\top \big] = n_j^{-1}\bSigma_j$. Similarly, for $\bS_{L,\bgamma}(\btheta)$, we have
$$\begin{aligned}
    \mathrm{E}\Big[ \bZ_j \big(y_j - \bW_j^\top \bbeta - \bZ_j^\top \bgamma \big)\Big|\bX_j,\bZ_j\Big] = \bZ_j \bX_j^\top \big(\bbeta_0 - \bbeta \big) + \bZ_j \bZ_j^\top \big(\bgamma_0 - \bgamma \big). 
\end{aligned}$$
Letting $(\bX,\bZ)$ denote an independent copy of $(\bX_j,\bZ_j)$, we have
$$
\begin{aligned}
& \mathrm{E}[\bS_{L,\bbeta}(\btheta)] = -2\,\mathrm{E}\big[\bX \bX^\top \big] \big(\bbeta_0 - \bbeta \big) 
 - 2\, \mathrm{E}\big[\bX \bZ^\top \big] \big(\bgamma_0 - \bgamma \big) \\
& \mathrm{E}[\bS_{L,\bgamma}(\btheta)] = -2\,\mathrm{E}\big[\bZ \bX^\top \big] \big(\bbeta_0 - \bbeta \big) 
 - 2\, \mathrm{E}\big[\bZ \bZ^\top \big] \big(\bgamma_0 - \bgamma \big).
 \end{aligned} 
$$
As a consequence of Lemma 1, $\bS_{L,\bbeta}(\btheta)$ and $\bS_{L,\bgamma}(\btheta)$ converge uniformly to $\mathrm{E}[\bS_{L,\bbeta}(\btheta)]$ and $\mathrm{E}[\bS_{L,\bgamma}(\btheta)]$, the first $p+q+1$ components of $\bS_0(\btheta)$. It is straightforward to see that the corresponding system of equations $$\mathrm{E}[\bS_{L,\bbeta}(\btheta)]=\bm{0}\qquad \mathrm{and}\qquad \mathrm{E}[\bS_{L,\bgamma}(\btheta)]= \bm{0}$$ has a unique solution at $\btheta = \btheta_0 =(\bbeta_0,\bgamma_0)$.
\subsection{Proof of Statement 2}
Consider next the phase function-based criterion $\tilde{D}(\btheta)$ directly. From Lemma 1, we have the uniform convergence of $\tilde{D}(\btheta)$ to a limiting function $D_0(\btheta)$. We will now evaluate this limiting function. Recall that
\begin{align}
\tilde{D}(\btheta) = &  \int_{0}^{t^*} \Bigg(
C_y(t) \left[\sum_{j=1}^{n} q_j \sin\left\{t\left(\bW_j^\top\bbeta + 
\bZ_j^\top\bgamma\right)\right\} \right]  \nonumber \\ 
&   - S_y(t) \left[\sum_{j=1}^{n} q_j\cos \left\{t\left(\bW_j^\top\bbeta +  \bZ_j^\top\bgamma\right)\right\} \right] \Bigg)^2 K_{t^*}(t)dt.
\end{align}
Let $V_0 = \bX^\top\bbeta_0 + \bZ^\top\bgamma_0$ and $Y = V_0 + \varepsilon$. For arbitrary $t$, by the weak law of large numbers,
$$C_y(t) = \dfrac{1}{n} \sum_{j=1}^{n} \cos(y_jt) \stackrel{p}{\rightarrow} \mathrm{E}\left[\cos(Yt) \right] = \text{Re}[\phi_{Y}(t)] = \text{Re}[\phi_{V_0}(t)] \phi_\varepsilon(t),
$$
where the last equality follows upon noting that $\varepsilon$ has a real-valued characteristic function. Similarly, $S_y(t) \stackrel{p}{\rightarrow} \text{Im}[\phi_{V_0}(t)] \phi_\varepsilon(t).
$
Furthermore, noting that for any $\btheta$, the random variables $\sin\big[t(\bW_j^\top \bbeta + \bZ_j^\top \bgamma)\big]$ and $\cos\big[t(\bW_j^\top \bbeta + \bZ_j^\top \bgamma)\big]$ are bounded, and subsequently have finite variances. By a generalized weak law of large numbers,
$$\sum_{j=1}^{n} q_j \sin\big\{t\left(\bW_j^\top\bbeta + 
\bZ_j^\top\bgamma\right)\big\}  \stackrel{p}{\rightarrow} \mathrm{E}\left[\sum_{j=1}^{n} q_j \sin\left\{t\left(\bW_j^\top\bbeta + 
\bZ_j^\top\bgamma\right)\right\} \right].$$
Letting $V_j(\bbeta,\bgamma) = \bX_j^\top \bbeta + \bZ_j^\top \bgamma$, we have
$$
\mathrm{E} \left[ \sin\left\{t\left(\bW_j^\top\bbeta + 
\bZ_j^\top\bgamma\right)\right\} \right] = \text{Im}\left\{\phi_{V_j(\bbeta,\bgamma)}(t) \right\} \phi_{\bU_j^\top\bbeta}(t), 
$$
where we make use of the fact that $\bU_j^\top\bbeta$ has a symmetric distribution about zero and hence a real-valued characteristic function. Since $(\bX_j,\bZ_j)$ are \textit{iid} by Condition C1, the random variables $V_j(\bbeta,\bgamma), ~j=1,\ldots, n$ are also \textit{iid} and have a common characteristic function $\phi_{V(\bbeta,\bgamma)}(t)=\phi_{V_j(\bbeta,\bgamma)}(t)$ for $j=1,\ldots,n$. As a consequence, we have
$$\mathrm{E}\left[\sum_{j=1}^{n} q_j \sin\left\{t\left(\bW_j^\top\bbeta + 
\bZ_j^\top\bgamma\right)\right\} \right] = \text{Im}\left\{\phi_{V(\bbeta,\bgamma)}(t) \right\} \left(\sum_{j=1}^{n} q_j \phi_{\bU_j^\top\bbeta}(t) \right).$$
Similarly,
$$\mathrm{E}\left[\sum_{j=1}^{n} q_j \cos\left\{t\left(\bW_j^\top\bbeta + 
\bZ_j^\top\bgamma\right)\right\} \right] = \text{Re}\left\{\phi_{V(\bbeta,\bgamma)}(t) \right\} \left(\sum_{j=1}^{n} q_j \phi_{\bU_j^\top\bbeta}(t) \right).$$
Letting $h(t,\bbeta) = \phi_{\varepsilon}(t)^2 \left[\sum_{j=1}^{n} q_j \phi_{\bU_j^\top\bbeta}(t) \right]^2$, and recalling the established uniform convergence for $(\bbeta,\bgamma)\in \bm{\Theta}$, the statistic $\tilde{D}(\btheta)$ converges uniformly to 
$$
\begin{aligned}
D_0(\btheta) = & \int_{0}^{t^*} \Big[ \text{Re}\left\{\phi_{V_0}\left(t\right)\right\} \text{Im}\left\{\phi_{V(\bbeta,\bgamma)}(t)\right\} \\
& - \text{Im}\left\{\phi_{V_0}(t)\right\} \text{Re}\left\{\phi_{V(\bbeta,\bgamma)}(t)\right\} \Big]^2 h(t,\bbeta) K_{t^*}(t) dt \\
& = \int_0^{t^\ast} \Big[\mathrm{Im}\{\phi_{V_0-V(\bbeta,\bgamma)}(t)\}\Big]^2 h(t,\bbeta) K_{t^\ast}dt,
\end{aligned}
$$
where random variables $V_0$ and $V(\bbeta,\bgamma)$ are independent. Note that $$\text{Im}\left\{\phi_{V_0 - V(\bbeta,\bgamma)}(t)\right\} = 0 \text{ for all } t\in\mathbb{R}$$ \textit{if and only if} the distribution of $V_0 - V(\bbeta,\bgamma)$ is symmetric about $0$. When Condition C4 holds, this is only true for $\btheta=\btheta_0 = (\bbeta_0,\bgamma_0)$. On the other hand, when Condition C4 does not hold, $D_0(\btheta)$ may have infinitely many global minima, but one of those minima still occurs at $\btheta = \btheta_0$. 

We conclude by noting that $\tilde{D}(\btheta)$ is continuous, as is the gradient vector $\nabla \tilde{D}(\btheta) = \partial \tilde{D}(\btheta) / \partial \btheta$. From Lemma 1, it subsequently follows that $\nabla \tilde{D}(\btheta)$ also converges uniformly to $\nabla D_0(\btheta)$. Thus, $\nabla D_0(\btheta_0)=\bm{0}$, even though this is not a unique solution to this system of equations. Note also that $\nabla D_0(\btheta_0)$ represents the last $p+q+1$ elements of $\bm{S}_0(\btheta)$. The proof is now complete.

\section{Calculating the Quasi-Likelihood Weights} \label{sec: QL weights}

The quasi-likelihood weights are defined in Section 3.2 of the main paper to be the minimizer of the $L_2$ discrepancy
$$
L(\bm{q}) = \sum_{j=1}^{n}({\bW}_j - \hat{\bm\mu}_q)^\top \left(\bm\Sigma_x + n_j^{-1} \bm\Sigma_j\right)^{-1}({\bW}_j - \hat{\bm\mu}_q)^\top,
$$
where $\hat{\bm\mu}_q = \sum_{j=1}^{n} q_j \bW_j$, subject to $q_j \geq 0$, $j=1,\ldots, n$ and $\sum_{j=1}^{n} q_j = 1$. To find the minimizer, let $\bm\Omega_j = \bm\Sigma_x + n_j^{-1} \bm\Sigma_j$ for $j=1,\ldots, n$. Some algebraic manipulation gives
\begin{align}
	L(\bm{q}) =& \sum_{j=1}^{n} \bW_j ^\top \bm\Omega_j^{-1} \bW_j - \hat{\bm\mu}_{q}^\top \left(\sum_{j=1}^{n} \bm\Omega_j^{-1} \bW_j\right) \nonumber \\
    & -\left(\sum_{j=1}^{n} \bW_j ^\top \bm\Omega_j^{-1}\right) \hat{\bm\mu}_{q}+ \hat{\bm\mu}_{q}^\top \left(\sum_{j=1}^{n} \bm\Omega_j^{-1}\right) \hat{\bm\mu}_{q}. \nonumber
\end{align}
Note that $L(\bm{q})$ is a function of $\bm{q}$ only through $\hat{\bm{\mu}}_q$. To calculate the weights, we define the function $D(\bm{q})$ to be only the terms in $L(\bm{q})$ involving $\hat{\bm{\mu}}_q$, and also introducing two Lagrange-multiplier type terms. The first of these ensures the weights $q_j$ sum to $1$, while the second ensures a numerically stable solution by constraining the squared differences between the $q_j$. The resulting function to be minimized is
\begin{eqnarray}
	D(\bm{q},\lambda) &=& -\hat{\bm\mu}_q^\top \left(\sum_{j=1}^{n} \bm\Omega_j^{-1} \bW_j\right)  - \left(\sum_{j=1}^{n} \bW_j ^\top \bm\Omega_j^{-1}\right) \hat{\bm\mu}_{q} + \hat{\bm\mu}_{q}^\top \left(\sum_{j=1}^{n} \bm\Omega_j^{-1}\right) \hat{\bm\mu}_{q} \nonumber \\
	&&  + 2\lambda \left(\sum_{j=1}^{n} q_j - 1\right) + \gamma \sum_{j,k}(q_j-q_k)^2. \nonumber
\end{eqnarray}
This function is minimized over $(\bm{q},\lambda)$, while $\gamma$ is a user-specified constant ensuring a numerically stable solution. Now, defining
\[\mathbf{A}_1 = \sum_{j=1}^{n} \bm\Omega_j^{-1} \bW_j \quad \text{and}\quad \mathbf{A}_2 = \sum_{j=1}^{n} \bm\Omega_j^{-1}\]
the target function can be written in the convenient form
\[
D(\bm{q},\lambda) = -2\hat{\bm\mu}_{q}^\top \mathbf{A}_1  + \hat{\bm\mu}_{q}^\top \mathbf{A}_2 \hat{\bm\mu}_{q} + 2\lambda \left(\sum_{j=1}^{n} q_j - 1\right) + \gamma \sum_{i,j}(q_i-q_j)^2. 
\]
This function is quadratic in $(\bm{q},\lambda)$ with a global minimum that can be found by solving a linear system of equations. Taking partial derivatives of with respect to the $q_k$, for $k=1,\ldots,n$, we have estimating equations
\[
\frac{\partial D}{\partial q_k} = - 2\bV_k^\top \mathbf{A}_1 + 2\bV_k^\top \mathbf{A}_2 \hat{\bm\mu}_{\bm{q}} + 2\lambda + 2\gamma\sum_{j\neq k} (q_k - q_j) = 0.
\]
Here, $\bV_k$ denotes a $n\times 1$ column vector of zeros, with the $k$th entry equal to $1$. Writing these estimating equations explicitly in terms of the weights $q$ gives
\[
-\bV_k^\top \mathbf{A}_1 + \sum_{j=1}^{n} q_j \left(\bV_k^\top \mathbf{A}_2 \bV_j \right) + \lambda + \gamma\sum_{j \neq k} (q_k - q_j)= 0,\ k=1,\ldots,n.\]
The minimizer of $D(\bm{q},\lambda)$ is found by solving the linear system
\[
 \resizebox{\textwidth}{!}{%
$
\begin{pmatrix}
	\bV_1^\top \mathbf{A}_2\bV_1+(n-1)\gamma & \bV_1^\top \mathbf{A}_2\bV_2-\gamma & \cdots & \bV_1^\top \mathbf{A}_2\bV_n-\gamma & 1  \\
	\bV_2^\top \mathbf{A}_2\bV_1-\gamma & \bV_2^\top \mathbf{A}_2\bV_2+(n-1)\gamma & \cdots & \bV_2^\top \mathbf{A}_2\bV_n-\gamma & 1  \\
	\vdots & \vdots & \ddots & \vdots & \vdots \\
	\bV_n^\top \mathbf{A}_2\bV_1-\gamma & \bV_n^\top \mathbf{A}_2X_2-\gamma & \cdots & \bV_n^\top \mathbf{A}_2\bV_n+(n-1)\gamma & 1 \\
	1 & 1 & \cdots & 1 & 0
\end{pmatrix}
\begin{pmatrix}
	q_1 \\ q_2 \\ \vdots \\ q_n \\ \lambda 
\end{pmatrix} 
= 
\begin{pmatrix}
	\bV_1^\top \mathbf{A}_1 \\ \bV_2^\top \mathbf{A}_1 \\ \vdots \\ \bV_n^\top \mathbf{A}_1 \\ 1
\end{pmatrix}.
$
}
\]

Numerical exploration suggests the solution $\hat{\bm{q}}$ of this system, which is easily obtained numerically, is fairly robust with regards to the specific choice of $\gamma$. In a simulation study not reported here, the choices $\gamma=1/n$ and $\gamma = \log(n)$ resulted in nearly identical estimators of the observation-specific weights. Furthermore, any value of $\gamma>0$ resulted in a numerically stable system to solve. We therefore use $\gamma = 1/n$ in the remainder of the paper.

\section{Additional Simulation Results} \label{sec: Additional Sims}
\subsection{Simulation for simple linear EIV models}

In this section, we report a set of simulation results for the simple errors-in-variables model $y_j = \gamma + \beta X_j + \epsilon_j, ~ W_{jk} = X_j + U_{jk}$ for $j=1,\ldots, n$ and $k=1,\ldots, n_{\text{rep}}$. In this setting, the true covariates $X_j$ are generated from the half-normal distribution scaled to have variance $1$, i.e. $X_j \stackrel{iid}{\sim} (1-2/\pi)^{-1/2}\, \big\vert N(0, 1) \big\vert$. Three scenarios are considered for the distribution of the measurement error terms, namely normal, $U_{jk} \sim N(0,\sigma_j^2)$, Student's t with $2.5$ degrees of freedom, $U_{jk} \sim \sigma_j/\sqrt{5}\, t_{2.5}$, and a contaminated normal,  $U_{jk} \sim \sigma_j/\sqrt{10.9}\{0.9 N(0, 1) + 0.1N(0, 10^2)\}$. In each scenario, the $U_{jk}$ are generated independently, have mean $0$, and are scaled to have $\mathrm{Var}(U_{jk}) = \sigma_j^2$ for $k=1,\ldots,n_{\mathrm{rep}}$. In all the three scenarios, the measurement error variances $\sigma_j^2$ are generated from the uniform distribution $n_\mathrm{rep} \times U(0.2, 1.5)$, so the signal-to-noise ratio $\mathrm{Var}(X_j)/\mathrm{Var}(U_j)$ for the averaged replicate measurement error $U_j$ ranges from $2/3$ (fairly weak signal) to $5$ (fairly strong signal). The true intercept and slope are set to be $\gamma_{00} = 2$ and $\beta = 1$, respectively. The regression error $\epsilon_j$'s are generated to match the distribution of the measurement error in each scenario and with constant variance $\sigma_\varepsilon^2 = 0.25$.

For each generated sample, we compute the same set of estimators and used the same criterion 
$\mathrm{det(\mathrm{MSE}_\mathrm{rob})}$ as described in Section 4.2 of the main paper.  Note that in the simple linear EIV model, the two heteroscedastic weighting schemes (minimax and quasi-likelihood) result in the same estimates.   
Table \ref{tab:table_setting_I} summarizes the results.  

\begin{table}[t]
\centering
\caption{Simple Linear EIV model performance naive OLS (Naive), moment-corrected (MC), and GMM estimators with equal (Equal), minimax (MM) and quasi-likelihood (QL) weights with $n_{\mathrm{rep}} \in \{2,3\}$ as measured by $\mathrm{det(1000 \times \mathrm{MSE}_\mathrm{rob})}$.}
\label{tab:table_setting_I}
\vspace{.4cm}
\resizebox{0.98\textwidth}{!}{\begin{tabular}{lr rrrr rrrr}
  \toprule[1.5pt]
$\mathbf{U}$ & $n$ & \multicolumn{4}{c}{$n_{\mathrm{rep}} = 2$} & \multicolumn{4}{c}{$n_{\mathrm{rep}} = 3$} \\[1pt] \cmidrule(lr){3-6} \cmidrule(lr){7-10}
& & Naive & MC & \multicolumn{2}{c}{GMM} & Naive & MC & \multicolumn{2}{c}{GMM}\\[2pt]
 & & & & Equal & QL & & & Equal & QL  \\[2pt]  
  \hline \addlinespace
Normal & 250 & 381.98 & 29.58 & 26.80 & 25.80 & 397.80 & 28.47 & 25.08 & 25.24 \\ 
   & 500 & 181.67 & 9.44 & 6.21 & 6.65 & 160.34 & 8.13 & 6.52 & 6.10 \\ 
   & 1000 & 88.78 & 2.13 & 1.63 & 1.56 & 90.83 & 1.80 & 1.40 & 1.40 \\ 
  $t_{2.5}$ & 250 & 263.14 & 20.89 & 10.56 & 8.74 & 322.03 & 17.25 & 9.32 & 7.99 \\ 
   & 500 & 128.75 & 5.66 & 2.29 & 1.90 & 143.20 & 3.75 & 2.05 & 1.60 \\ 
   & 1000 & 87.26 & 1.44 & 0.52 & 0.42 & 72.70 & 1.27 & 0.49 & 0.36 \\ 
  Cont.Normal & 250 & 19.96 & 22.19 & 2.18 & 1.97 & 20.05 & 12.25 & 1.94 & 1.26 \\ 
   & 500 & 44.30 & 15.28 & 0.98 & 0.66 & 46.25 & 11.49 & 0.90 & 0.48 \\ 
   & 1000 & 98.12 & 14.27 & 0.75 & 0.48 & 90.87 & 12.58 & 0.66 & 0.37 \\ \bottomrule[1.5pt]
\end{tabular}}
\end{table}

It comes as no surprise that the naive estimator has the worst performance among the estimators considered. The large values observed demonstrate the consequences of ignoring measurement errors when they are present. When considering the various corrected estimators, the GEE estimators outperform the moment-corrected approach regardless of the weighting scheme used. The improvement of GMM estimators over moment correction is especially pronounced in contaminated normal error scenario, with the $t_{2.5}$ error scenario also showing some large decreases in estimation error. These cases illustrate the value of combining moment correction with the phase function-based approach. When comparing the two GMM weighting schemes, the estimator with quasi-likelihood weights also outperforms the equal weights estimator in almost all cases. The only exceptions occur under the normal measurement error model, where equal weighting in two instances performs better than quasi-likelihood weighting.

\subsection{Additional results for multiple linear EIV models}
In this section, we summarize simulation results equivalent to those in Section 4.2 of the main paper. Tables S2 through S4 are analogous to Tables 1 through 3, but with $n_{\mathrm{rep}}=3$ whereas the main paper presents results for $n_{\mathrm{rep}}=2$. For greater specificity regarding the simulation configurations, see the descriptions in Section 4.2 of the main paper. We note here here that similar conclusions can be drawn for the case with $n_{\mathrm{rep}}=3$ replicates. Specifically, when the measurement error follows a $t_{2.5}$ or contaminated normal distribution, the GMM estimators outperform moment correction. On the other hand, when the measurement error follows a normal distribution, there isn't a clear preference for either the moment corrected or GMM estimators, each in turn outperforming the other. Finally, Table S5 is analogous to Table 4 of the main paper, showing the accuracy with which the standard error is estimated for the model parameters.

\begin{table}[ht]
\centering
\caption{Setting I performance of uncontaminated OLS (True), naive OLS (Naive), moment-corrected (MC), and GMM estimators with equal (Equal), minimax (MM) and quasi-likelihood (QL) weights with $n_{\mathrm{rep}} = 3$ as measured by $\mathrm{det(1000 \times \mathrm{MSE}_\mathrm{rob})}$.}
\vspace{.4cm}
\label{tab:table_setting_II_3replicates}
\begin{tabular}{llrrrrrrr}
  \hline
  \toprule[1.5pt]
$\mathbf{R}$ & $\mathbf{U}$ & $n$ & True & Naive & MC & \multicolumn{3}{c}{GMM} \\  
& & & & & & Equal & MM & QL \\[2pt] 
  \hline
  \addlinespace
 $\rho=0$ & Normal & 250 & 0.010 & 14.237 & 1.565 & 2.082 & 2.023 & 2.087 \\ 
   &  & 500 & 0.001 & 3.385 & 0.188 & 0.182 & 0.180 & 0.180 \\ 
   &  & 1000 & 0.000 & 0.944 & 0.028 & 0.029 & 0.031 & 0.030 \\ 
   & $t_{2.5}$ & 250 & 0.012 & 20.132 & 1.827 & 0.967 & 0.805 & 0.802 \\ 
   &  & 500 & 0.001 & 4.778 & 0.108 & 0.075 & 0.072 & 0.071 \\ 
   &  & 1000 & 0.000 & 1.364 & 0.018 & 0.014 & 0.012 & 0.012 \\ 
   & Cont.Normal & 250 & 0.009 & 13.100 & 1.009 & 0.853 & 0.650 & 0.676 \\ 
   &  & 500 & 0.001 & 3.744 & 0.128 & 0.100 & 0.073 & 0.080 \\ 
   &  & 1000 & 0.000 & 1.036 & 0.017 & 0.015 & 0.012 & 0.013 \\[.5em] 
  $\rho=0.5$ & Normal & 250 & 0.009 & 34.094 & 2.511 & 3.097 & 2.892 & 2.771 \\ 
   &  & 500 & 0.001 & 6.323 & 0.286 & 0.252 & 0.274 & 0.253 \\ 
   &  & 1000 & 0.000 & 1.715 & 0.042 & 0.042 & 0.044 & 0.042 \\ 
   & $t_{2.5}$ & 250 & 0.008 & 39.974 & 2.152 & 1.134 & 1.004 & 1.129 \\ 
   &  & 500 & 0.001 & 8.377 & 0.166 & 0.105 & 0.108 & 0.111 \\ 
   &  & 1000 & 0.000 & 3.272 & 0.027 & 0.020 & 0.017 & 0.019 \\ 
   & Cont.Normal & 250 & 0.009 & 41.149 & 2.271 & 1.761 & 1.400 & 1.475 \\ 
   &  & 500 & 0.001 & 10.698 & 0.311 & 0.241 & 0.157 & 0.191 \\ 
   &  & 1000 & 0.000 & 2.413 & 0.035 & 0.027 & 0.019 & 0.023 \\ 
  \bottomrule[1.5pt]
\end{tabular}
\end{table}

\begin{table}[ht]
\centering
\caption{Setting II performance of uncontaminated OLS (True), naive OLS (Naive), moment-corrected (MC), and GMM estimators with equal (Equal), minimax (MM) and quasi-likelihood (QL) weights with $n_{\mathrm{rep}} = 3$ as measured by $\mathrm{det(1000 \times \mathrm{MSE}_\mathrm{rob})}$.}
\label{tab:table_setting_III_3replicates}
\begin{tabular}{llrrrrrrr}
  \hline
  \toprule[1.5pt]
$\mathbf{R}$ & $\mathbf{U}$ & $n$ & True & Naive & MC & \multicolumn{3}{c}{GMM} \\  
& & & & & & Equal & MM & QL \\[2pt] 
  \hline
  \addlinespace
$\rho=0$ & Normal & 250 & 0.001 & 34.108 & 2.509 & 2.777 & 2.673 & 2.403 \\ 
   &  & 500 & 0.000 & 1.716 & 0.096 & 0.108 & 0.122 & 0.108 \\ 
   &  & 1000 & 0.000 & 0.089 & 0.002 & 0.003 & 0.003 & 0.003 \\ 
   & $t_{2.5}$ & 250 & 0.002 & 13.080 & 1.388 & 1.133 & 0.814 & 0.746 \\ 
   &  & 500 & 0.000 & 2.394 & 0.051 & 0.042 & 0.036 & 0.034 \\ 
   &  & 1000 & 0.000 & 0.255 & 0.002 & 0.001 & 0.002 & 0.001 \\ 
   & Cont.Normal & 250 & 0.001 & 18.461 & 1.571 & 1.483 & 1.460 & 1.041 \\ 
   &  & 500 & 0.000 & 1.495 & 0.055 & 0.055 & 0.055 & 0.036 \\ 
   &  & 1000 & 0.000 & 0.057 & 0.001 & 0.001 & 0.001 & 0.001 \\[.5em] 
  $\rho=0.5$ & Normal & 250 & 0.001 & 51.039 & 9.331 & 14.770 & 18.866 & 14.313 \\ 
   &  & 500 & 0.000 & 4.915 & 0.234 & 0.266 & 0.291 & 0.257 \\ 
   &  & 1000 & 0.000 & 0.336 & 0.009 & 0.010 & 0.011 & 0.011 \\ 
   & $t_{2.5}$ & 250 & 0.002 & 92.843 & 6.001 & 3.647 & 2.989 & 2.726 \\ 
   &  & 500 & 0.000 & 5.910 & 0.143 & 0.105 & 0.104 & 0.090 \\ 
   &  & 1000 & 0.000 & 0.648 & 0.006 & 0.004 & 0.003 & 0.003 \\ 
   & Cont.Normal & 250 & 0.001 & 60.581 & 4.318 & 3.805 & 3.496 & 2.731 \\ 
   &  & 500 & 0.000 & 4.936 & 0.150 & 0.149 & 0.128 & 0.094 \\ 
   &  & 1000 & 0.000 & 0.296 & 0.005 & 0.004 & 0.004 & 0.003 \\
  \bottomrule[1.5pt]
\end{tabular}
\end{table}

\begin{table}[ht]
\centering
\caption{Setting III performance of uncontaminated OLS (True), naive OLS (Naive), moment-corrected (MC), and GMM estimators with equal (Equal), minimax (MM) and quasi-likelihood (QL) weights with $n_{\mathrm{rep}} = 3$ as measured by $\mathrm{det(1000 \times \mathrm{MSE}_\mathrm{rob})}$.}
\vspace{.4cm}
\label{tab:table_setting_IV_3replicates}
\begin{tabular}{llrrrrrrr}
  \hline
  \toprule[1.5pt]
$\mathbf{R}$ & $\mathbf{U}$ & $n$ & True & Naive & MC & \multicolumn{3}{c}{GMM} \\  
& & & & & & Equal & MM & QL \\[2pt] 
  \hline
  \addlinespace
$\rho=0$ & Normal & 250 & 0.001 & 14.862 & 1.997 & 1.972 & 1.964 & 1.976 \\ 
   &  & 500 & 0.000 & 0.789 & 0.043 & 0.043 & 0.043 & 0.043 \\ 
   &  & 1000 & 0.000 & 0.042 & 0.001 & 0.001 & 0.001 & 0.001 \\ 
   & $t_{2.5}$ & 250 & 0.001 & 13.456 & 1.132 & 1.087 & 0.978 & 1.056 \\ 
   &  & 500 & 0.000 & 0.837 & 0.022 & 0.022 & 0.020 & 0.020 \\ 
   &  & 1000 & 0.000 & 0.118 & 0.001 & 0.001 & 0.001 & 0.001 \\ 
   & Cont.Normal & 250 & 0.001 & 21.877 & 2.346 & 2.379 & 2.159 & 2.282 \\ 
   &  & 500 & 0.000 & 1.116 & 0.053 & 0.053 & 0.049 & 0.052 \\ 
   &  & 1000 & 0.000 & 0.055 & 0.002 & 0.002 & 0.001 & 0.001 \\[.5em] 
  $\rho=0.5$ & Normal & 250 & 0.001 & 51.724 & 4.557 & 4.554 & 4.543 & 4.555 \\ 
   &  & 500 & 0.000 & 3.059 & 0.145 & 0.145 & 0.145 & 0.145 \\ 
   &  & 1000 & 0.000 & 0.180 & 0.005 & 0.005 & 0.005 & 0.005 \\ 
   & $t_{2.5}$ & 250 & 0.001 & 32.805 & 2.819 & 2.777 & 2.470 & 2.635 \\ 
   &  & 500 & 0.000 & 2.937 & 0.081 & 0.076 & 0.069 & 0.070 \\ 
   &  & 1000 & 0.000 & 0.354 & 0.003 & 0.003 & 0.003 & 0.003 \\ 
   & Cont.Normal & 250 & 0.001 & 44.184 & 4.592 & 4.594 & 4.243 & 4.462 \\ 
   &  & 500 & 0.000 & 3.465 & 0.149 & 0.149 & 0.135 & 0.145 \\ 
   &  & 1000 & 0.000 & 0.205 & 0.004 & 0.005 & 0.004 & 0.004 \\
   \bottomrule[1.5pt]
\end{tabular}
\end{table}

\begin{table}[ht]
\centering
\caption{Monte Carlo standard errors (MC-SE) and average of average of the bootstrap plug-in standard errors (Avg-SE) for the GMM estimators with the minimax weighting scheme in simulation settings I and III with $n_\text{rep}=3$ replicates and $\rho = 0.5$.}
\begin{tabular}{lllrrrr}
  \toprule[1.5pt]
Setting & $\bU$ & Coeff & \multicolumn{2}{c}{$n=500$} & \multicolumn{2}{c}{$n=1000$} \\
 & & & MC-SE & Avg-SE & MC-SE & Avg-SE\\ 
  \hline
II & Normal & $\hat\beta_1$ & 0.030 & 0.029 & 0.021 & 0.021 \\ 
   &  & $\hat\beta_2$ & 0.028 & 0.029 & 0.022 & 0.021 \\ 
   &  & $\hat\gamma_{00}$ & 0.043 & 0.044 & 0.031 & 0.031 \\ 
   & $t_{2.5}$ & $\hat\beta_1$ & 0.023 & 0.026 & 0.022 & 0.019 \\ 
   &  & $\hat\beta_2$ & 0.027 & 0.025 & 0.019 & 0.019 \\ 
   &  & $\hat\gamma_{00}$ & 0.035 & 0.036 & 0.027 & 0.025 \\[1em]
   
  IV & Normal & $\hat\beta_1$ & 0.036 & 0.033 & 0.021 & 0.022 \\ 
   &  & $\hat\beta_2$ & 0.032 & 0.030 & 0.021 & 0.022 \\ 
   &  & $\hat\gamma_{00}$& 0.061 & 0.049 & 0.037 & 0.035 \\ 
   &  & $\hat\gamma_{1}$ & 0.026 & 0.027 & 0.020 & 0.019 \\ 
   &  & $\hat\gamma_{2}$ & 0.028 & 0.028 & 0.019 & 0.019 \\ 
   & $t_{2.5}$ & $\hat\beta_1$ & 0.029 & 0.032 & 0.020 & 0.021 \\ 
   &  & $\hat\beta_2$ & 0.027 & 0.027 & 0.020 & 0.020 \\ 
   &  & $\hat\gamma_{00}$ & 0.043 & 0.046 & 0.032 & 0.030 \\ 
   &  & $\hat\gamma_{1}$ & 0.028 & 0.026 & 0.022 & 0.018 \\ 
   &  & $\hat\gamma_{2}$ & 0.029 & 0.025 & 0.016 & 0.018 \\ 
   \hline
\end{tabular}
\end{table}

\FloatBarrier

\FloatBarrier

\begin{spacing}{1.6}

\bibliographystyle{apalike}
\bibliography{citation}

\begin{thebibliography}{}

\bibitem[Buonaccorsi, 2010]{buonaccorsi2010measurement}
Buonaccorsi, J.~P. (2010).
\newblock {\em Measurement Error: Models, Methods, and Applications}.
\newblock CRC press.

\bibitem[Carroll et~al., 2006]{carroll2006measurement}
Carroll, R.~J., Ruppert, D., Stefanski, L.~A., and Crainiceanu, C.~M. (2006).
\newblock {\em Measurement Error in Nonlinear Models: A Modern Perspective}.
\newblock Chapman and Hall/CRC.

\bibitem[Carroll and Stefanski, 1990]{carroll1990approximate}
Carroll, R.~J. and Stefanski, L.~A. (1990).
\newblock Approximate quasi-likelihood estimation in models with surrogate
  predictors.
\newblock {\em Journal of the American Statistical Association},
  85(411):652--663.

\bibitem[Delaigle and Hall, 2016]{delaigle2016methodology}
Delaigle, A. and Hall, P. (2016).
\newblock Methodology for non-parametric deconvolution when the error
  distribution is unknown.
\newblock {\em Journal of the Royal Statistical Society: Series B: Statistical
  Methodology}, pages 231--252.

\bibitem[Erickson et~al., 2014]{erickson2014minimum}
Erickson, T., Jiang, C.~H., and Whited, T.~M. (2014).
\newblock Minimum distance estimation of the errors-in-variables model using
  linear cumulant equations.
\newblock {\em Journal of Econometrics}, 183(2):211--221.

\bibitem[Gillard, 2014]{gillard2014method}
Gillard, J. (2014).
\newblock Method of moments estimation in linear regression with errors in both
  variables.
\newblock {\em Communications in Statistics-Theory and Methods},
  43(15):3208--3222.

\bibitem[Hanfelt and Liang, 1997]{hanfelt1997approximate}
Hanfelt, J.~J. and Liang, K.-Y. (1997).
\newblock Approximate likelihoods for generalized linear errors-in-variables
  models.
\newblock {\em Journal of the Royal Statistical Society: Series B (Statistical
  Methodology)}, 59(3):627--637.

\bibitem[Hansen, 1982]{hansen1982large}
Hansen, L.~P. (1982).
\newblock Large sample properties of generalized method of moments estimators.
\newblock {\em Econometrica: Journal of the Econometric Society}, pages
  1029--1054.

\bibitem[Hansen et~al., 1996]{hansen1996finite}
Hansen, L.~P., Heaton, J., and Yaron, A. (1996).
\newblock Finite-sample properties of some alternative gmm estimators.
\newblock {\em Journal of Business \& Economic Statistics}, 14(3):262--280.

\bibitem[Higdon and Schafer, 2001]{higdon2001maximum}
Higdon, R. and Schafer, D.~W. (2001).
\newblock Maximum likelihood computations for regression with measurement
  error.
\newblock {\em Computational Statistics \& Data Analysis}, 35(3):283--299.

\bibitem[Hoadley, 1971]{hoadley1971asymptotic}
Hoadley, B. (1971).
\newblock Asymptotic properties of maximum likelihood estimators for the
  independent not identically distributed case.
\newblock {\em The Annals of Mathematical Statistics}, pages 1977--1991.

\bibitem[Hu and Kalbfleisch, 1997]{hu1997estimating}
Hu, F. and Kalbfleisch, J.~D. (1997).
\newblock Estimating equations and the bootstrap.
\newblock {\em Lecture Notes-Monograph Series}, pages 405--416.

\bibitem[Newey and McFadden, 1994]{newey1994chapter}
Newey, W.~K. and McFadden, D. (1994).
\newblock Chapter 36: Large sample estimation and hypothesis testing.
\newblock {\em Handbook of Econometrics}, 4:2111--2245.

\bibitem[Nghiem and Potgieter, 2018]{nghiem2018density}
Nghiem, L. and Potgieter, C.~J. (2018).
\newblock Density estimation in the presence of heteroscedastic measurement
  error of unknown type using phase function deconvolution.
\newblock {\em Statistics in Medicine}, 37(25):3679--3692.

\bibitem[Nghiem et~al., 2020]{nghiem2020estimation}
Nghiem, L.~H., Byrd, M.~C., and Potgieter, C.~J. (2020).
\newblock Estimation in linear errors-in-variables models with unknown error
  distribution.
\newblock {\em Biometrika}, 107(4):841--856.

\bibitem[P{\"o}tscher and Prucha, 1989]{potscher1989uniform}
P{\"o}tscher, B.~M. and Prucha, I.~R. (1989).
\newblock A uniform law of large numbers for dependent and heterogeneous data
  processes.
\newblock {\em Econometrica: Journal of the Econometric Society}, pages
  675--683.

\bibitem[Reiers{\o}l, 1941]{reiersol1941confluence}
Reiers{\o}l, O. (1941).
\newblock Confluence analysis by means of lag moments and other methods of
  confluence analysis.
\newblock {\em Econometrica: Journal of the Econometric Society}, pages 1--24.

\bibitem[Rousseeuw and Driessen, 1999]{rousseeuw1999fast}
Rousseeuw, P.~J. and Driessen, K.~V. (1999).
\newblock A fast algorithm for the minimum covariance determinant estimator.
\newblock {\em Technometrics}, 41(3):212--223.

\bibitem[Song, 2021]{song2021nonparametric}
Song, W. (2021).
\newblock Nonparametric inference methods for berkson errors.
\newblock In {\em Handbook of Measurement Error Models}, pages 271--292.
  Chapman and Hall/CRC.

\bibitem[Stefanski and Carroll, 1987]{stefanski1987conditional}
Stefanski, L.~A. and Carroll, R.~J. (1987).
\newblock Conditional scores and optimal scores for generalized linear
  measurement-error models.
\newblock {\em Biometrika}, 74(4):703--716.

\bibitem[Stefanski and Cook, 1995]{stefanski1995simulation}
Stefanski, L.~A. and Cook, J.~R. (1995).
\newblock Simulation-extrapolation: the measurement error jackknife.
\newblock {\em Journal of the American Statistical Association},
  90(432):1247--1256.

\bibitem[Wald, 1940]{wald1940fitting}
Wald, A. (1940).
\newblock The fitting of straight lines if both variables are subject to error.
\newblock {\em The Annals of Mathematical Statistics}, 11(3):284--300.

\bibitem[Xue-Kun~Song, 2000]{xue2000multivariate}
Xue-Kun~Song, P. (2000).
\newblock Multivariate dispersion models generated from gaussian copula.
\newblock {\em Scandinavian Journal of Statistics}, 27(2):305--320.

\end{thebibliography}

\end{spacing}
\end{document}